\documentclass[twocolumn]{aastex63}
\usepackage{graphicx}
\usepackage{enumerate}
\usepackage{amssymb, amsmath}
\usepackage{natbib}
\usepackage{color}
\usepackage{ulem}
\usepackage{dblfnote}
\usepackage{appendix}
\usepackage{hyperref}
\usepackage[stable]{footmisc}
\usepackage{relsize}

\definecolor{red}{rgb}{1.0,0.0,0.0}
\def\pasa{PASA}
\def\procspie{Proc.~SPIE}

\begin{document}

\title{SCExAO/CHARIS Direct Imaging Discovery of a 20 au Separation, Low-Mass Ratio Brown Dwarf Companion to an Accelerating Sun-like Star\footnote{Based in part on data collected at Subaru Telescope, which is operated by the National Astronomical Observatory of Japan.}}

\email{currie@naoj.org}
\author{Thayne Currie}
\affiliation{Subaru Telescope, National Astronomical Observatory of Japan, 
650 North A`oh$\bar{o}$k$\bar{u}$ Place, Hilo, HI  96720, USA}
\affiliation{NASA-Ames Research Center, Moffett Blvd., Moffett Field, CA, USA}
\affiliation{Eureka Scientific, 2452 Delmer Street Suite 100, Oakland, CA, USA}
\author{Timothy D. Brandt}
\affiliation{Department of Physics, University of California, Santa Barbara, Santa Barbara, California, USA}
\author{Masayuki Kuzuhara}
\affiliation{Astrobiology Center of NINS, 2-21-1, Osawa, Mitaka, Tokyo, 181-8588, Japan}
\affiliation{National Astronomical Observatory of Japan, 2-21-2, Osawa, Mitaka, Tokyo 181-8588, Japan}
\author{Jeffrey Chilcote}
\affiliation{Department of Physics, University of Notre Dame, South Bend, IN, USA}
\author{Olivier Guyon}
\affiliation{Subaru Telescope, National Astronomical Observatory of Japan, 
650 North A`oh$\bar{o}$k$\bar{u}$ Place, Hilo, HI  96720, USA}
\affil{Astrobiology Center of NINS, 2-21-1, Osawa, Mitaka, Tokyo, 181-8588, Japan}
\affil{Steward Observatory, The University of Arizona, Tucson, AZ 85721, USA}
\affil{College of Optical Sciences, University of Arizona, Tucson, AZ 85721, USA}
\author{Christian Marois}
\affiliation{National Research Council of Canada Herzberg, 5071 West Saanich Rd, Victoria, BC, V9E 2E7, Canada}
\affiliation{Department of Physics and Astronomy,University of Victoria, 3800 Finnerty Rd, Victoria, BC, V8P 5C2, Canada}
\author{Tyler D. Groff}
\affiliation{NASA-Goddard Space Flight Center, Greenbelt, MD, USA}
\author{Julien Lozi}
\affiliation{Subaru Telescope, National Astronomical Observatory of Japan, 
650 North A`oh$\bar{o}$k$\bar{u}$ Place, Hilo, HI  96720, USA}
\author{Sebastien Vievard}
\affiliation{Subaru Telescope, National Astronomical Observatory of Japan, 
650 North A`oh$\bar{o}$k$\bar{u}$ Place, Hilo, HI  96720, USA}
\author{Ananya Sahoo}
\affiliation{Subaru Telescope, National Astronomical Observatory of Japan, 
650 North A`oh$\bar{o}$k$\bar{u}$ Place, Hilo, HI  96720, USA}
\author{Vincent Deo}
\affiliation{Subaru Telescope, National Astronomical Observatory of Japan, 
650 North A`oh$\bar{o}$k$\bar{u}$ Place, Hilo, HI  96720, USA}
\author{Nemanja Jovanovic}
\affiliation{Department of Astronomy, California Institute of Technology, 1200 E. California Blvd.,Pasadena, CA, 91125, USA}
\author{Frantz Martinache}
\affiliation{Universit\'{e} C\^{o}te d'Azur, Observatoire de la C\^{o}te d'Azur, CNRS, Laboratoire Lagrange, France}
\author{Kevin Wagner}
\affil{Steward Observatory, The University of Arizona, Tucson, AZ 85721, USA}
\affil{NASA Hubble/Sagan Fellow, NASA Nexus for Exoplanet System Science, Earths in Other Solar Systems Team}
\author{Trent Dupuy}
\affiliation{Institute for Astronomy, University of Edinburgh, Blackford Hill, Edinburgh EH9 3HJ, UK ; Centre for Exoplanet Science, University of Edinburgh, Edinburgh, UK}
\author{Matthew Wahl}
\affiliation{Subaru Telescope, National Astronomical Observatory of Japan, 
650 North A`oh$\bar{o}$k$\bar{u}$ Place, Hilo, HI  96720, USA}
\author{Michael Letawsky}
\affiliation{Subaru Telescope, National Astronomical Observatory of Japan, 
650 North A`oh$\bar{o}$k$\bar{u}$ Place, Hilo, HI  96720, USA}
\author{Yiting Li}
\affiliation{Department of Physics, University of California, Santa Barbara, Santa Barbara, California, USA}
\author{Yunlin Zeng}
\affiliation{School of Physics, Georgia Institute of Technology, 837 State Street, Atlanta, Georgia, USA}
\author{G. Mirek Brandt}
\affiliation{Department of Physics, University of California, Santa Barbara, Santa Barbara, California, USA}
\author{Daniel Michalik}
\affiliation{European Space Agency (ESA/ESTEC): Noordwijk, South Holland, NL}

\author{Carol Grady}
\affiliation{Eureka Scientific, 2452 Delmer Street Suite 100, Oakland, CA, USA}
\author{Markus Janson}
\affiliation{Department of Astronomy, Stockholm University, AlbaNova University Center, 10691, Stockholm, Sweden}
\author{Gillian R. Knapp}
\affiliation{Department of Astrophysical Sciences, Princeton University, Princeton, NJ USA}
\author{Jungmi Kwon}
\affiliation{ISAS/JAXA, 3-1-1 Yoshinodai, Chuo-ku, Sagamihara, Kanagawa 252-5210, Japan}
\author{Kellen Lawson}
\affiliation{Homer L. Dodge Department of Physics, University of Oklahoma, Norman, OK 73071, USA}
\author{Michael W. McElwain}
\affiliation{NASA-Goddard Space Flight Center, Greenbelt, MD, USA}
\author{Taichi Uyama}
\affiliation{Infrared Processing and Analysis Center, California Institute of Technology, Pasadena, CA 91125, USA}
\author{John Wisniewski}
\affiliation{Homer L. Dodge Department of Physics, University of Oklahoma, Norman, OK 73071, USA}
\author{Motohide Tamura}
\affil{Astrobiology Center of NINS, 2-21-1, Osawa, Mitaka, Tokyo, 181-8588, Japan}
\affiliation{National Astronomical Observatory of Japan, 2-21-2, Osawa, Mitaka, Tokyo 181-8588, Japan}
\affiliation{Department of Astronomy, Graduate School of Science, The University of Tokyo, 7-3-1, Hongo, Bunkyo-ku, Tokyo, 113-0033, Japan}


\shortauthors{Currie et al.}
\begin{abstract}
We present the direct imaging discovery of a substellar companion to the nearby Sun-like star, HD 33632 Aa, at a projected separation of $\sim$ 20 au, obtained with SCExAO/CHARIS integral field spectroscopy complemented by Keck/NIRC2 thermal infrared imaging.   The companion, HD 33632 Ab, induces a 10.5$\sigma$ astrometric acceleration on the star as detected with the {\it Gaia} and {\it Hipparcos} satellites.  SCExAO/CHARIS $JHK$ (1.1--2.4 $\mu$m) spectra and Keck/NIRC2 $L_{\rm p}$ (3.78 $\mu$m) photometry are best matched by a field L/T transition object: an older, higher gravity, and less dusty counterpart to HR 8799 cde.
Combining our astrometry with {\it Gaia/Hipparcos} data and archival Lick Observatory radial-velocities, we measure a dynamical mass of 46.4 $\pm$ 8 $M_{\rm J}$ and an eccentricity of $e$ $<$0.46 at 95\% confidence.
HD 33632 Ab's mass and mass ratio (4.0\% $\pm$ 0.7\%) are comparable to the low-mass brown dwarf GJ 758 B and intermediate between the more massive brown dwarf HD 19467 B and the (near-)planet mass companions to HR 2562 and GJ 504.  
Using {\it Gaia} to select for direct imaging observations with the newest extreme adaptive optics systems can reveal substellar or even planet-mass companions on solar system-like scales at an increased frequency compared to blind surveys.
\end{abstract}

\section{Introduction}
 Facility and now extreme adaptive optics systems have provided the first direct detections of exoplanets around nearby, young stars
 \citep[][]{Marois2008a,Marois2010,Kuzuhara2013,Currie2014a,Currie2015,Macintosh2015,Keppler2018}. 
 \textit{Blind} direct imaging surveys have provided the first constraints on the frequency of jovian planets and low-mass brown dwarfs at orbital distances of 10--500 au \citep[][]{Brandt2014,Nielsen2019}.

Unfortunately, the low yield of blind direct imaging surveys shows that exoplanets detectable with current instruments are rare \citep[e.g.][]{Nielsen2019}, especially around solar and subsolar-mass stars.   
The few companions found by these surveys typically exceed five jovian masses and orbit well beyond separations where the jovian planet frequency peaks \citep[][]{Fernandes2019}.  Improving blind survey yields will only be possible with substantial contrast gains at small separations and/or sensitivity gains at wider separations \citep[][]{Crepp2011}.  Limited sample sizes and sparse coverage of ages, temperatures, and surface gravities impede our understanding of the atmospheric evolution of gas giant planets.  

Targeted searches focused on stars showing evidence of gravitational pulls from massive planets could improve survey yields.  While radial-velocity data have been used for sample selection with some success for finding (sub-)stellar companions \citep[][]{Crepp2014}, astrometry can also select direct imaging targets, even around stars whose activity and spectral type preclude precise radial-velocity measurements.     The {\it Hipparcos}-{\it Gaia} Catalog of Accelerations \citep[HGCA;][]{Brandt2018} provides absolute astrometry for 115,000 nearby stars, including those with clear dynamical evidence for unseen massive companions.   HGCA-derived accelerations can provide dynamical masses of imaged exoplanets and low-mass brown dwarfs independent of luminosity evolution models and irrespective of stellar age uncertainties \citep{Brandt2019,Dupuy2019}. 

In this Letter, we report the direct imaging discovery of a low-mass substellar companion 20 au from the nearby Sun-like star HD 33632 Aa using the Subaru Coronagraphic Extreme Adaptive Optics Project \citep[SCExAO,][]{Jovanovic2015} coupled to the CHARIS integral field spectrograph \citep{Groff2016} and using the NIRC2 camera on Keck.  The HGCA identifies an astrometric acceleration for the primary induced by HD 33632 Ab, providing a dynamical mass of $46 \pm 8$ $M_{\rm J}$ for the companion.  CHARIS spectra and NIRC2 photometry reveal HD 33632 Ab to be a higher-gravity, higher-mass counterpart to young L/T transition planet-mass objects such as HR 8799 cde with a low eccentricity more typical of directly-imaged planets than brown dwarfs.


\begin{deluxetable*}{llllllllll}
     \tablewidth{0pt}
    \tablecaption{HD 33632 Observing Log\label{obslog}}
    \tablehead{\colhead{UT Date} & \colhead{Instrument} &  \colhead{Seeing$^{c}$ (\arcsec{})} &{Filter} & \colhead{$\lambda$ ($\mu m$)$^{a}$} 
    & \colhead{$t_{\rm exp}$} & \colhead{$N_{\rm exp}$} & \colhead{$\Delta$PA ($^{o}$)} & \colhead{Reduction Strategy} & SNR$^{d}$ (HD 33632 Ab)}
    \startdata
    20181018 & SCExAO/CHARIS$^{a}$ &  0.5 & $JHK$ & 1.16--2.37& 45.7 & 14 & 3.0 & SDI & 22\\
    20181101 & Keck/NIRC2 &  0.7 & $L_{\rm p}$ &3.78 & 50 & 26 & 13.9 & ADI & 11\\
    20200831 & SCExAO/CHARIS$^{a}$ &  0.6 & $JHK$ & 1.16--2.37& 31 & 86 & 17.8 & ADI, ADI+SDI & 41\\
    20200901 & SCExAO/CHARIS$^{a,b}$ &  0.5--1.0 & $JHK$ & 1.16--2.37& 45.7 & 26 & 20.8 & ADI, ADI+SDI & 27 \\
    \enddata
    \tablecomments{
    a) For CHARIS data, this column refers to the wavelength range.  For NIRC2 imaging data, it refers to the central wavelength. b) Twenty six frames ($\sim$~20 minutes) of the 31 August 2020 data were also obtained with the star dithered by 1\farcs{}1 to the north and east. c) From the Canada France Hawaii Telescope seeing monitor.  d) The SNR ratio of HD 33632 Ab in the wavelength-collapsed image reduced using ``conservative" ADI/A-LOCI algorithm settings or SDI/A-LOCI for the case of 2018 October SCExAO/CHARIS data.   
    }
    \label{obslog_hd33632}
    \end{deluxetable*}
   
   \begin{figure*}
    \centering
       \vspace{-0.4in}
    \includegraphics[width=0.95\textwidth,trim=0mm 0mm 0mm 0mm,clip]{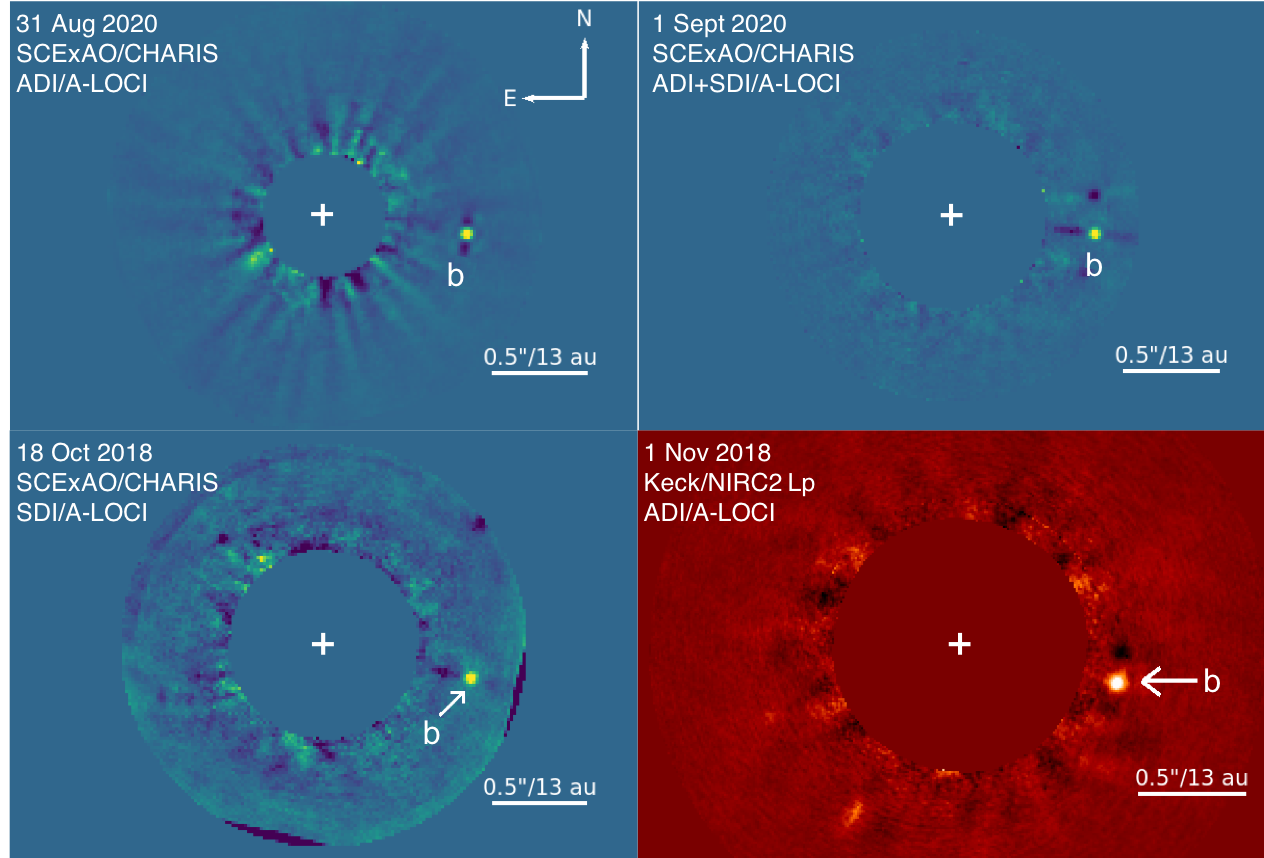}
    \vspace{-0.1in}
    \caption{HD 33632 Ab detections in 2020 (top panels) and 2018 (bottom panels) reduced using A-LOCI with different combinations of ADI, SDI, and ADI+SDI.   Our adopted rotation gap/radial scaling criteria precludes access to masked regions. }
    \vspace{-0.in}
    \label{fig:images}
\end{figure*} 

\section{Systems Properties, Observations and Data}
The HD 33632 primary is a 26.56 $pc$ distant F8V 1.1 $M_{\odot}$ main sequence star \citep{Takeda2007,GAIA2018}.  The star has an M star companion, 2MASS J05131845+3720463, at $\rho$ $\sim$ 34\arcsec{} (900 au) \citep{Scholz2016} with the same parallax as HD 33632 Aa and a proper motion difference of $\sim$3~mas\,yr$^{-1}$ \citep{GAIA2018,Brandt2018}\footnote{The Washington Double Star Catalog lists the system as WDS J05133+3720A and WDS J05133+3720B: a third interloping star is mislabeled as WDS J05133+3720C.  To avoid confusion, we describe the primary as HD 33632 Aa and the companion as HD 33632 Ab.}. 

Age estimates for the HD 33632 primary drawn from CaHK measurements (log($R'_{\rm HK}$) = $-$4.76 to $-$4.93) encompass $\sim$ 1.2--4.5 Gyr \citep[e.g.][]{Takeda2007,Pace2013}.  
The \citet{Mamajek2008} activity-rotation-age relation and gyrochronology as implemented in \citet{Brandt2014} favor the low end of this range ($\sim$~1-2.5 $Gyr$).   HD 33632 Aa's fractional X-ray luminosity (log($L_{\rm X}/L_{\rm bol}$ = $-$5.2) overlaps with those for 800 Myr-old Hyades stars \citep{Brandt2015, Freund2020}.   Li abundances are consistent with the Hyades but marginally consistent with 4 Gyr-old M67 \citep{Ramirez2012,Castro2016}.   Ages estimated using neutron capture elements abundances individually (e.g.~[Zr/Fe]) or in aggregate (e.g.~the average abundances of ``heavy" elements Ce, Ba, and La; [Y/Mg]) cluster around 1.5--2.5 $Gyr$ \citep[][]{Spina2018}.   Given activity, gyrochronology, and neutron capture abundance-derived estimates, we adopt a broad age range of 1.5$^{+3.0}_{-0.7}$ Gyr with a preference for 1.0--2.5 Gyr.

HGCA reveals the HD 33632 primary to have a statistically significant acceleration ($\chi^{2}$ $\approx$ 116) for a model of constant proper motion, equivalent to 10.5\,$\sigma$ with two degrees of freedom: evidence for the dynamical pull of a companion.  The acceleration is almost 1000 times larger than that expected from the M-dwarf companion.
The Lick radial-velocity survey monitored HD~33632 Aa for eleven years (18 Jan 1998 to 1 Feb 2009) without finding evidence of a planet or brown dwarf \citep{Fischer2014}: the companion must be 
widely separated.   Following estimates in \citet{Brandt2019}, its minimum mass at angles accessible by SCExAO/CHARIS ($\rho$ $\sim$ 0\farcs{}1--1\farcs{}05) ranges from a few jovian masses to over 50 $M_{\rm J}$.

Table \ref{obslog_hd33632} summarizes our observations with SCExAO/CHARIS and Keck/NIRC2 in 2018 and 2020.
The seeing ranged between $\theta_{\rm V}$ = 0\farcs{}4 and 1\farcs{}0; conditions were photometric each night.    
SCExAO achieved a high-fidelity AO correction;
we utilized the CHARIS integral field spectrograph \citep{Groff2016} in low spectral resolution (broadband) mode  covering the $JHK$ passbands simultaneously (1.16-2.37 $\mu$m at $\mathcal{R}$ $\sim$ 18).   NIRC2 data were taken in the $L_{\rm p}$ broadband filter ($\lambda_{o}$ = 3.78 $\mu m$) using Keck's facility AO system.

All CHARIS data utilized satellite spots for precise astrometric and spectrophotometric calibration \citep[][]{Jovanovic2015-astrogrids}.  The CHARIS data were taken with a Lyot coronagraph (0\farcs{}139 radius occulting mask); the NIRC2 data were taken behind a Lyot coronagraph with a 0\farcs{}2 radius occulting mask.    All observations were conducted in pupil tracking mode, enabling angular differential imaging \citep[ADI;][]{Marois2006}; the CHARIS data also enable spectral differential imaging \citep[SDI;][]{Marois2000}. 

We extracted CHARIS data cubes using the standard CHARIS pipeline \citep{Brandt2017} and perfomed basic reduction steps including 
sky subtraction, image registration, and spectrophotometric calibration as in \citet{Currie2018}.   For spectrophotometric calibration, we adopted a Kurucz stellar atmosphere model appropriate for an F8V star. We reduced the NIRC2 data using the ADI-based pipeline from \cite{Currie2011}.     

For all data, we used A-LOCI for point-spread function (PSF) subtraction \citep{Currie2018}.
Due to a lack of field rotation, we subtracted the PSF for the 2018 CHARIS data in SDI-mode only.  The 2020 CHARIS data enabled PSF subtraction with ADI, SDI, and ADI+SDI (i.e.~performing SDI on the A-LOCI PSF subtraction residuals produced from ADI).   
In all cases, we used a large rotation gap/radial scaling criterion of $\delta$ $>$ 1 $\lambda$/D to limit self-subtraction. For the CHARIS SDI reduction (component), we employed a pixel mask over the subtraction zone \citep{Marois2010,Currie2015} to significantly reduce biasing of a companion spectrum.

Figure \ref{fig:images} shows the detection of a faint point source, hereafter HD 33632 Ab, $\rho$ $\approx$ 0\farcs{}75 west from the primary in each epoch.  The detection's signal-to-noise ratios (SNR) in conservative ADI/A-LOCI or SDI/A-LOCI reductions range between 11--22 in the 2018 discovery epoch to 41 for the 31 August 2020 CHARIS broadband follow-up data.  
CHARIS data do not identify any other companion within $\rho$ $\sim$ 1\farcs{}05.
In addition to resolving HD 33632 Ab, NIRC2 data identify a wider separation object at [E,N]\arcsec{} = [$-$1\farcs{}22, $-$1\farcs{}85] and $L_{\rm p}$ $\sim$ 16.3 (not shown).  However, 2020 CHARIS data reveal this object to be a background star based on its astrometry and its near-infrared brightness.

\section{Analysis}

\begin{figure*}
    \includegraphics[width=0.5\textwidth,trim=25mm 6mm 10mm 0mm,clip]{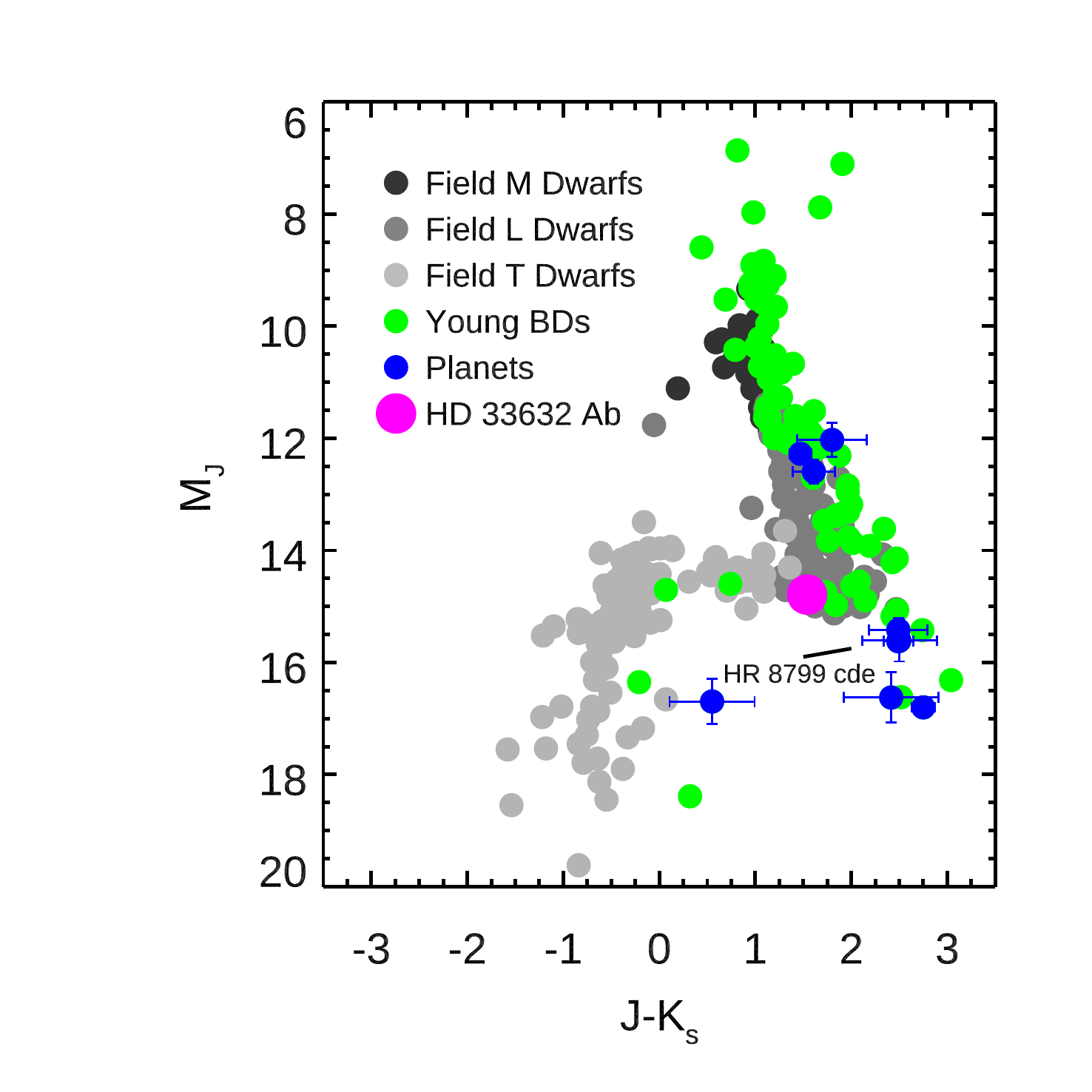}
    \includegraphics[width=0.5\textwidth,trim=25mm 6mm 10mm 0mm,clip]{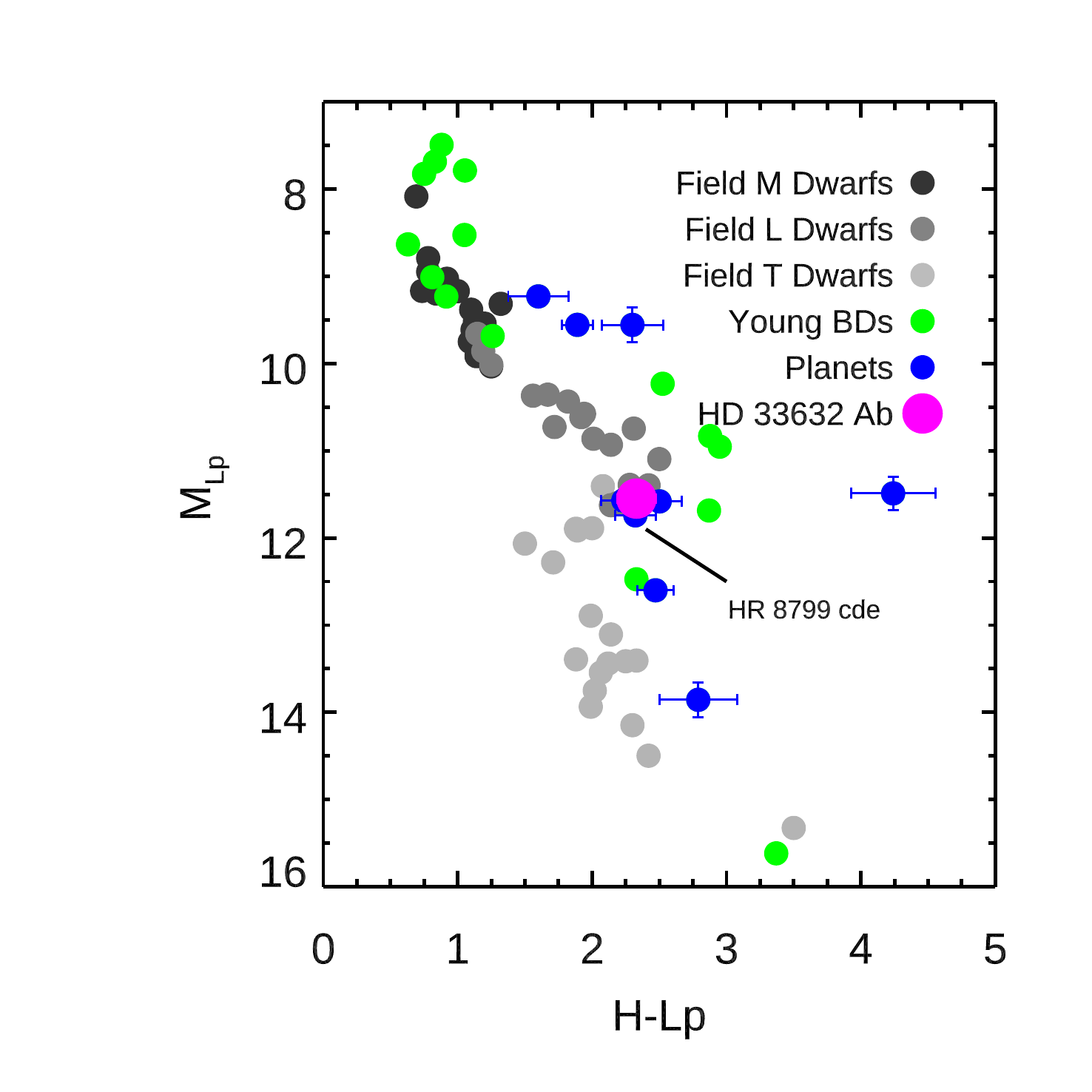}\\
     \includegraphics[width=0.5\textwidth,trim=15mm 4mm 8mm 10mm,clip]{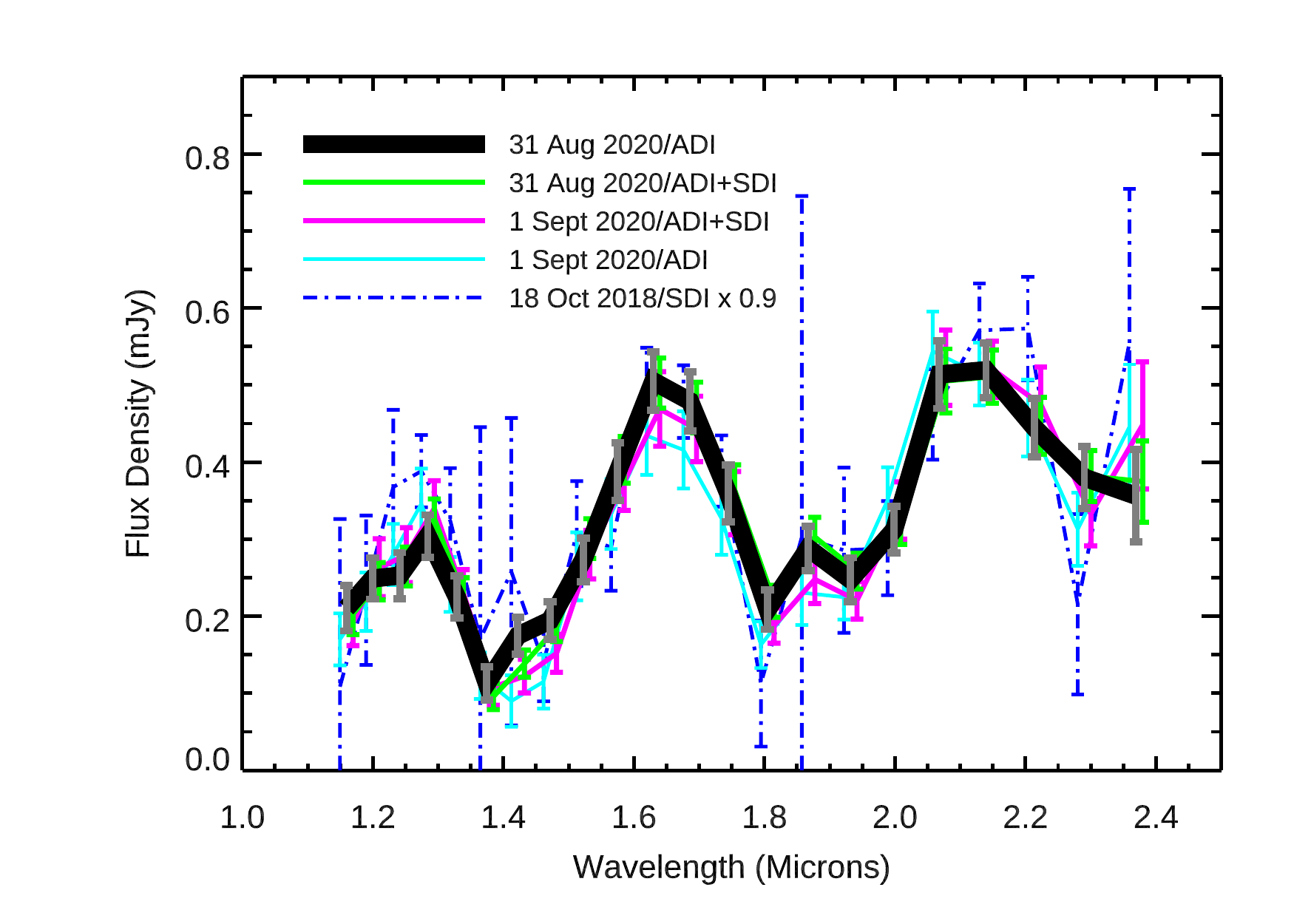}  
     \includegraphics[width=0.5025\textwidth,trim=20mm 6mm 10mm 10mm,clip]{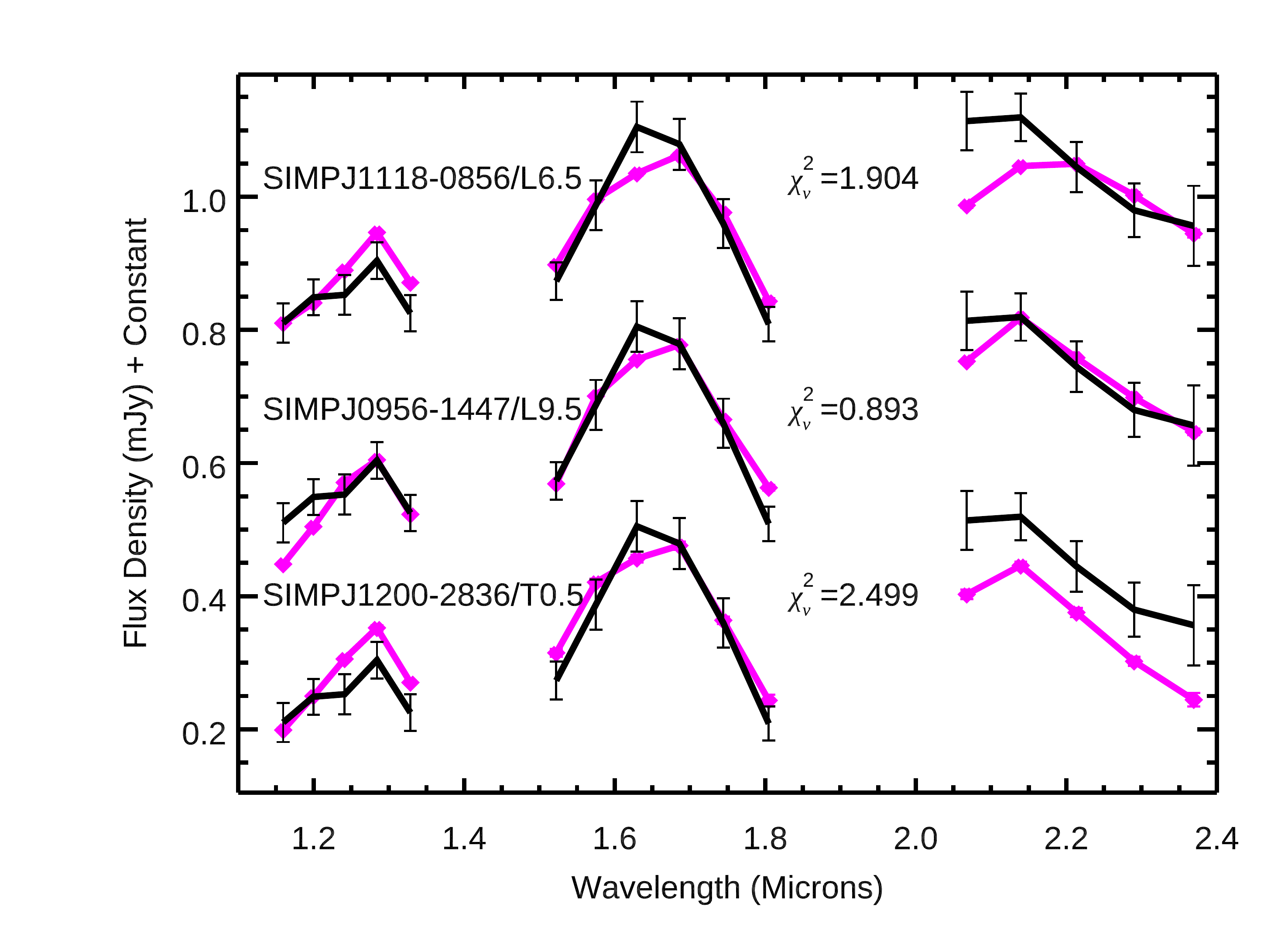}
     \vspace{-0.25in}
    \caption{(Top) $J$/$J$-$K_{\rm s}$ and $L_{\rm p}$/$H$-$L_{\rm p}$ color-magnitude diagram comparing field/young low mass MLT type (sub-)stellar objects, directly imaged exoplanets, and HD 33632 Ab.   Data draw from \citet{DupuyLiu2012} with updated photometry for directly-imaged planets: e.g. HR 8799 bcde \citep[drawn from][]{Zurlo2016,Currie2014b}.   
    (Bottom) (left) CHARIS spectrum extracted from 31 August 2020 data reduced with ADI/A-LOCI (black thick line, gray error bars) compared to spectra extracted on different nights or with different processing.  (right) The CHARIS HD 33632 Ab spectrum (black) compared to field brown dwarf spectra (magenta) near the L/T transition from the Montreal Spectral Library (binned to CHARIS's resolution).}
    \label{fig: empcomp}
\end{figure*}

\subsection{Infrared Colors, Spectrum, and Atmosphere of HD 33632 A\lowercase{b}}
To derive throughput-corrected spectrophotometry, we forward-modeled point sources at HD 33632 Ab's locations using stored LOCI coefficients \citep{Currie2018}.   We focus on ADI-only reductions due to the more straightforward applicability of ADI-reduced data to forward-modeling \citep[][]{Pueyo2016}.  We adopt our highest-quality data set (31 August 2020) for our analyses.  Our results agree with those from other data sets and with SDI reductions.  

We extract spectra from the best-fit centroid position of the wavelength-collapsed image.  Throughputs typically exceed 75\%; because we use ADI only, they are independent of HD 33632 Ab's spectrum.
HD 33632 Ab's broadband photometry in standard Mauna Kea Observatory filters is $J = 16.91 \pm 0.11$, $H = 16.00 \pm 0.09$, $K_{\rm s} = 15.37 \pm 0.09$, and $L_{\rm p} = 13.67 \pm 0.15$ ($\Delta\ignorespaces J,H,K_{\rm s} = 11.52, 10.83, 10.21$).   Photometry obtained at different epochs or with different reductions agrees with these results to within errors.

HD 33632 Ab lies at the transition from cloudy L dwarfs to (nearly) cloud free T dwarfs \citep[][]{Burrows2006}.  Figure \ref{fig: empcomp} places it on an infrared color-magnitude diagram.  The companion's $J$/$J$-$K_{\rm s}$ position is consistent with field L/T objects, while its $L_{\rm p}$/$H$-$L_{\rm p}$ position agrees with both field objects and the low gravity, cloudy exoplanets HR 8799 cde \citep{Currie2011,Barman2015}.


The CHARIS spectrum of HD 33632 Ab (Figure \ref{fig: empcomp}, bottom-left panel; Table \ref{spectrum_hd33632}) is highly peaked, characteristic of a substellar atmosphere shaped by water opacity.  The 2018 CHARIS spectrum is significantly noisier and offset by $\sim$ 10\% compared to the 2020 epoch data.  Otherwise, spectra extracted using different reduction techniques and taken on different epochs agree to within errors.   The detection significance of HD 33632 Ab reaches SNR = 20--30 in some channels for the 31 August 2020/ADI-reduced spectrum.   However, contributions from uncertainties in the absolute flux calibration (i.e. the satellite spot flux uncertainty) limit the spectrophotometric precision to about 10\%. 

We compared HD 33632 Ab's CHARIS spectrum to objects in the Montreal Spectral Library\footnote{\url{https://jgagneastro.com/the-montreal-spectral-library/}} \citep[e.g.][]{Gagne2014}, considering the impact of spatially and spectrally correlated noise \citep{GrecoBrandt2016}.   The spectral covariance at HD 33632 Ab's angular separation in the 31 Aug 2020 ADI/A-LOCI reduced data is nearly zero for off-diagonal elements 
indicating largely uncorrelated noise.   

The L9.5 field brown dwarfs SIMPJ0956-1447 and SIMPJ0150+3827 provide the best fit to HD 33632 Ab's spectrum (Figure \ref{fig: empcomp}, bottom-right panel).   While the Montreal library contains few objects later than L5, the $\chi^{2}$ distribution for library objects exhibits a local mininum between L6.5 and T0.5.   Thus, conservatively we assign HD 33632 Ab a spectral type of L9.5$^{+1.0}_{-3.0}$.   Adopting the mapping between spectral type and temperature from \citet{Stephens2009}, the object's effective temperature is 1300$^{+100}_{-100}$ K: slighty higher than that derived for the inner three HR 8799 planets \citep{Currie2011,Barman2015}.   Adopting the polynomial fit from \citet{Golimowski2004} and a distance of 26.56 $pc$, HD 33632 Ab's luminosity is $\log_{10}(L/L_{\rm \odot}) = -4.62^{+0.04}_{-0.08}$.   


\subsection{Astrometry and Common Proper Motion} 

As our forward-modeling revealed negligible astrometric bias and different reductions yielded consistent astrometry, we used gaussian centroiding to determine the position of HD 33632 Ab.   HD 33632 Ab is located at [E,N]$\arcsec{}$ = [$-$0\farcs{}761,$-$0\farcs{}176] $\pm$ [0\farcs{}005,0\farcs{}004] and [$-$0\farcs{}753,$-$0\farcs{}178] $\pm$ [0\farcs{}005,0\farcs{}005] in the 2018 CHARIS and NIRC2 data, respectively: the same within errors.   Our errors consider the centroiding precision and uncertainties in the pixel scale and true north\footnote{We reassessed the pixel scale and north position angle estimate for the 2020 data, following \citet{Currie2018} in using contemporaneous Keck/NIRC2 astrometry for HD 1160 B \citep{Nielsen2012} taken on 09 July 2020.   Adopting a north position angle of $-2.\!\!^\circ2$ and pixel scale of 0\farcs{}0162 as in \citet{Currie2018}, HD 1160 B's positions with NIRC2 and CHARIS agree to within 4 mas, well within errors.}.   

In the 2020 CHARIS data, HD 33632 Ab's position is [E,N]$\arcsec{}$ = [$-$0\farcs{}740,$-$0\farcs{}095] $\pm$ [0\farcs{}005,0\farcs{}003].   Between October 2018 and August 2020, a background star would be offset to the northeast by $\sim$ [0\farcs{}24,0\farcs{}18], inconsistent with HD 33632 Ab's observed displacement of [0\farcs{}021,0\farcs{}081].


  \begin{figure*}
    \begin{flushright}
    \includegraphics[width=1.0\textwidth]{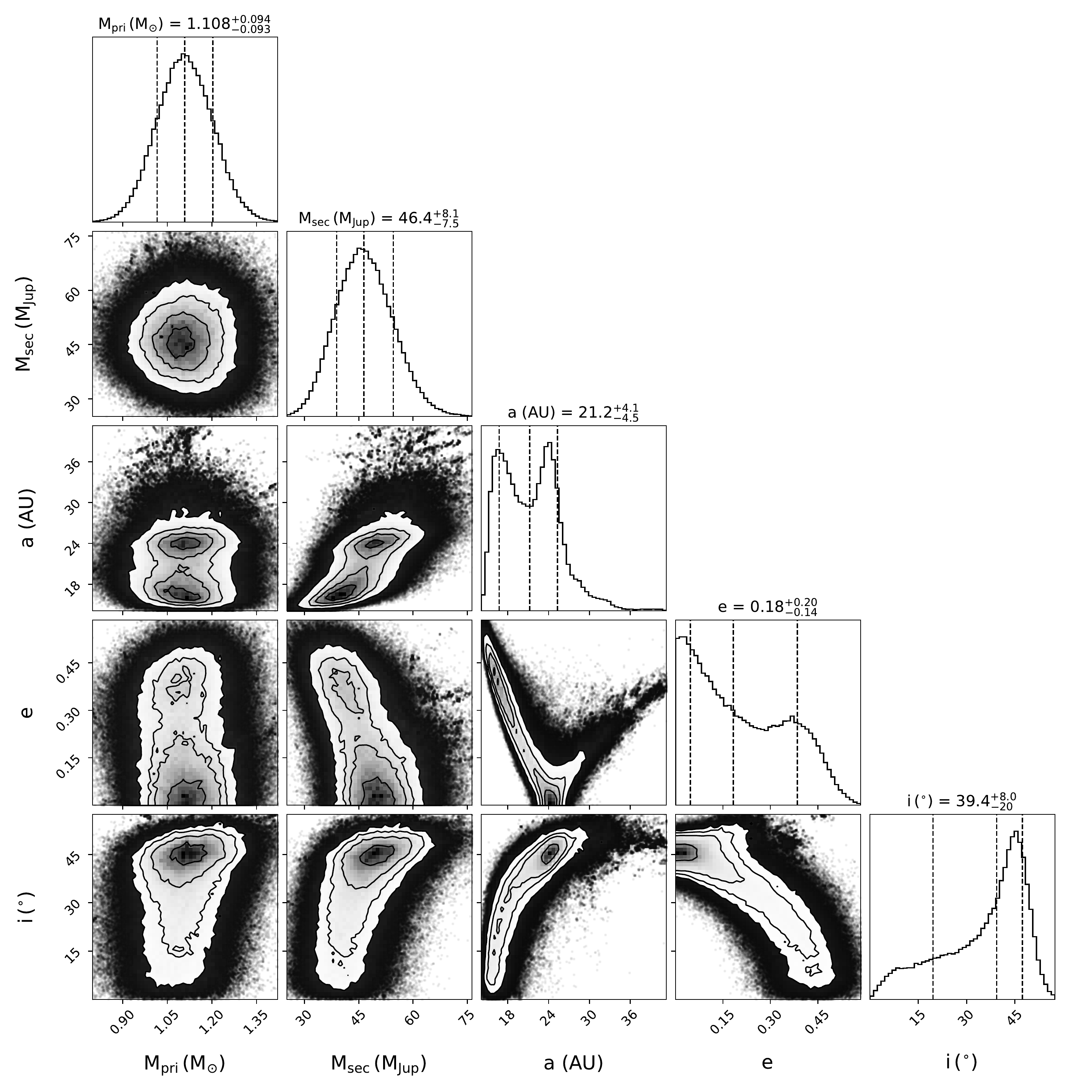} \\
    \vspace{-1.05\textwidth}
    \includegraphics[width=0.44\textwidth]{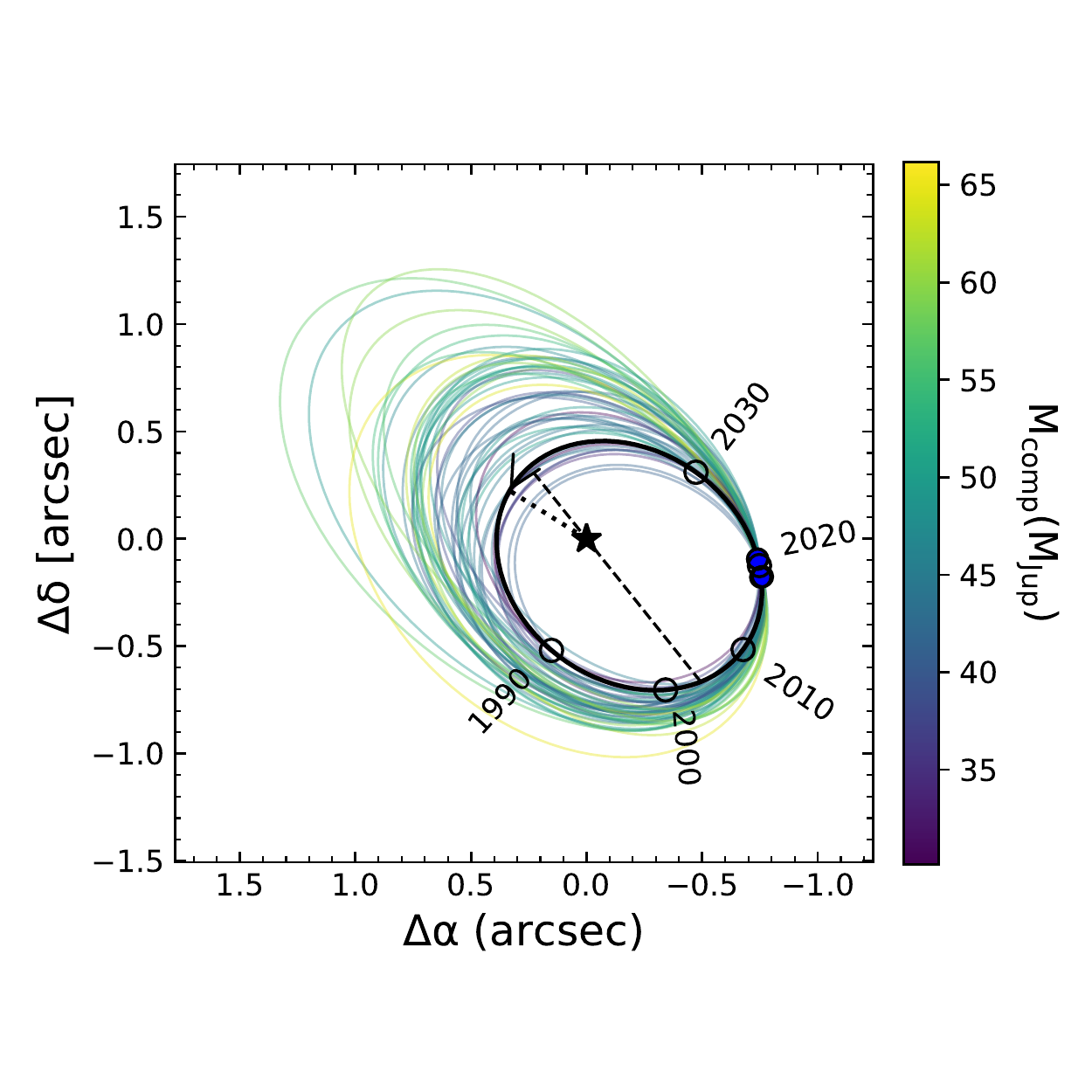} \\
    \vspace{-0.04\textwidth}
    \includegraphics[width=0.36\textwidth]{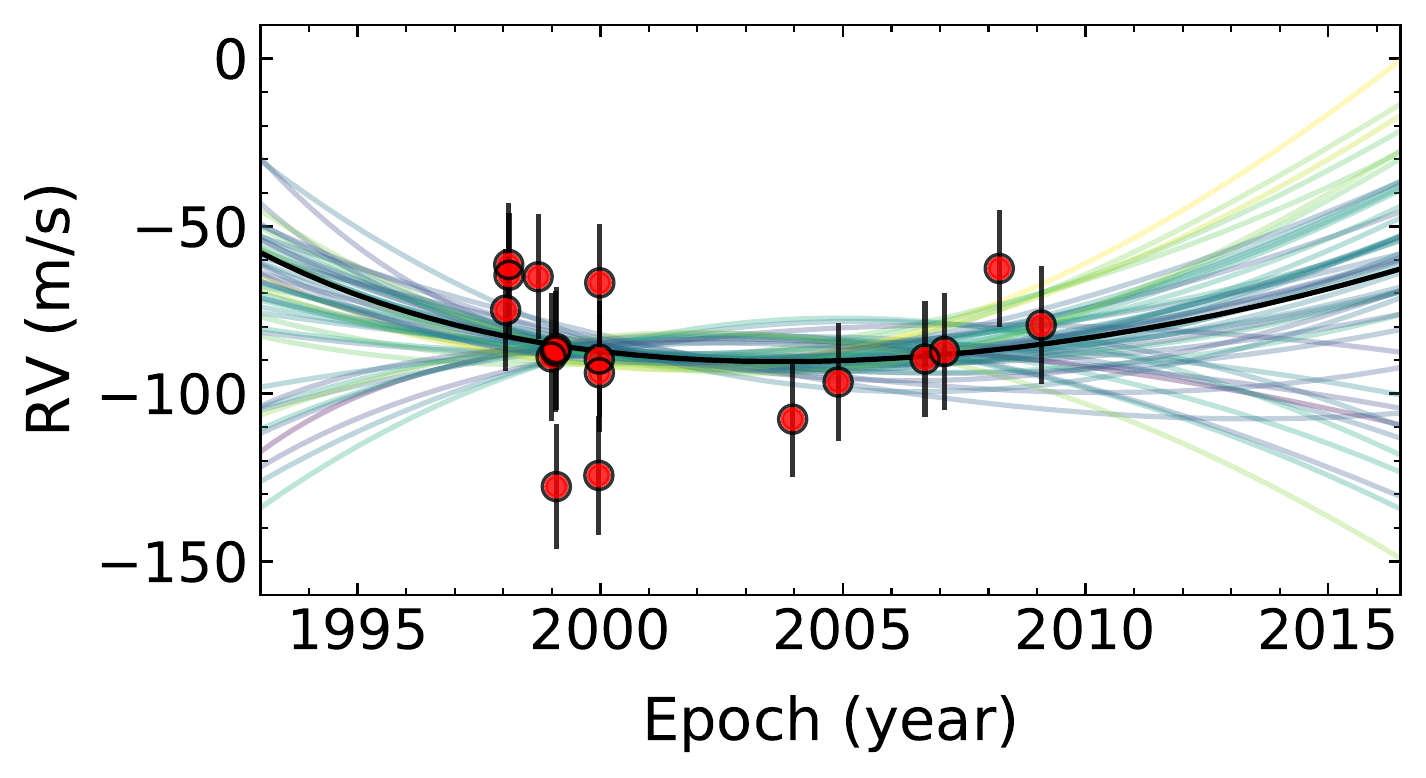}
    \vspace{0.43\textwidth}
    \end{flushright}
    \caption{Corner plot showing posterior distributions of selected orbital parameters.  The orbit fits used 
    {\it Hipparcos} and {\it Gaia} absolute astrometry, Lick Observatory precision radial-velocity measurements for HD 33632 Aa, and relative astrometry of HD 33632 Ab.  The posterior on HD 33632 Aa's mass is dominated by our adopted prior of $1.1 \pm 0.1~M_\odot$.   The insets display the best-fit orbit along with 100 random orbits drawn from our MCMC in both relative astrometry and radial velocity, color-coded by HD 33632 Ab's mass.  Zero radial velocity corresponds to the system barycenter of the best-fit orbit (black curve); all radial velocity data points and curves are shifted by the best-fit radial velocity offsets to align with the black curve.  }
    \vspace{-0.in}
    \label{fig:orbits}
\end{figure*} 

\begin{deluxetable}{llll}
     \tablewidth{0pt}
    \tablecaption{HD 33632 Ab Spectrum}
    \tablehead{\colhead{Wavelength ($\mu$m)} & \colhead{$F_{\rm \nu}$ (mJy)} &  \colhead{$\sigma$~$F_{\rm \nu}$ (mJy)} & \colhead{SNR}}
    \startdata
1.160 & 0.210 & 0.029 &   8.9 \\
1.200 & 0.249 & 0.027 &  12.6 \\
1.241 & 0.253 & 0.030 &  10.7 \\
1.284 & 0.304 & 0.028 &  13.9 \\
1.329 & 0.225 & 0.027 &  10.7 \\
1.375 & 0.113 & 0.022 &   5.5 \\
1.422 & 0.175 & 0.024 &   8.1 \\
1.471 & 0.195 & 0.024 &   9.3 \\
1.522 & 0.273 & 0.028 &  13.0 \\
1.575 & 0.387 & 0.037 &  14.4 \\
1.630 & 0.505 & 0.038 &  21.3 \\
1.686 & 0.479 & 0.038 &  20.8 \\
1.744 & 0.360 & 0.037 &  16.1 \\
1.805 & 0.209 & 0.026 &  10.2 \\
1.867 & 0.288 & 0.029 &  13.0 \\
1.932 & 0.247 & 0.028 &  10.9 \\
1.999 & 0.312 & 0.030 &  14.2 \\
2.068 & 0.514 & 0.044 &  28.0 \\
2.139 & 0.519 & 0.035 &  30.1 \\
2.213 & 0.445 & 0.038 &  22.7 \\
2.290 & 0.380 & 0.040 &  12.7 \\
2.369 & 0.356 & 0.060 &   6.5 \\
    \enddata
    \tablecomments{Throughput-corrected HD 33632 Ab spectrum extracted from 31 August 2020 data, reduced using ADI/A-LOCI.
    }
    \label{spectrum_hd33632}
    \end{deluxetable}

\subsection{Orbit and Dynamical Mass}

We use the open-source code {\tt orvara} (Brandt et al., submitted) to fit HD 33632 Ab's mass and orbit using a combination of HGCA measurements, Lick Observatory radial-velocity data \citep{Fischer2014}, and relative astrometry from CHARIS and NIRC2.  No other bound companion is detected with CHARIS, Lick data do not identify an inner companion, and the only other object seen with NIRC2 is a background star.   Given its separation, the M dwarf companion contributes $<$1\% of the acceleration seen by the HGCA.  Therefore, we assume that HD 33632 Ab is solely responsible for the HGCA acceleration.  The astrometric acceleration of $\sim$0.1~mas\,yr$^{-2}$ is more than 1000 times smaller than the apparent acceleration due to parallactic motion: it negligibly affects the {\it Gaia} parallax.  We fit for a radial velocity jitter, the six Keplerian orbital elements, and the masses of both components.  We adopt a prior on HD~33632~Aa's mass of $1.1\pm0.1~M_\odot$ and a log-uniform prior on the mass of HD~33632~Ab.  We analytically marginalize out the nuisance parameters of barycenter proper motion, parallax, and radial velocity zero point.

\begin{deluxetable*}{lccr}
\tablecaption{MCMC Orbit Fitting Priors and Results}
\tablewidth{0pt}
  \tablehead{
    Parameter &
    16/50/84\% quantiles &
    95\% confidence interval &
    Prior}
  \startdata
\multicolumn{4}{c}{Fitted Parameters} \\ \hline
RV jitter (m/s)  &       ${2.53}_{-0.18}^{+0.19}$ & (2.19, 2.91) & log-flat \\
$M_{\rm pri}$ ($M_\odot$)   &      ${1.108}_{-0.093}^{+0.094}$ & (0.93, 1.29) & Gaussian, $1.1 \pm 0.1$ \\
$M_{\rm sec}$ ($M_{\rm Jup}$)  &       ${46.4}_{-7.5}^{+8.1}$ & (32.5, 62.8) & $1/M_{\rm sec}$ (log-flat) \\
Semimajor axis $a$ (AU)    &     ${21.2}_{-4.5}^{+4.1}$ & (15.2, 30.9) & $1/a$  (log-flat) \\
$\sqrt{e} \sin \omega$\tablenotemark{\rm *}    &     ${0.05}_{-0.24}^{+0.21}$ & ($-$0.39, 0.47) & uniform \\
$\sqrt{e} \cos \omega$\tablenotemark{\rm *}  &       ${0.15}_{-0.52}^{+0.37}$ & ($-$0.62, 0.64) & uniform  \\
Inclination ($^\circ$)    &     ${39.4}_{-20}^{+8.0}$   & (6.0, 52.2) & $\sin i$ (geometric) \\
PA of the ascending node $\Omega$ ($^\circ$)  &      ${38.2}_{-7.0}^{+7.2}$ & (21.3, 256) & uniform \\
Mean longitude at 2010.0 ($^\circ$) &        ${21}_{-10}^{+14}$ & (10.3, 286) & uniform \\
Parallax (mas)   &      ${37.647}\pm 0.071$ & (37.52, 37.78) & Gaussian, $37.646 \pm 0.064$ \\
Barycenter $\mu_{\alpha*}$ (mas/yr) & $-144.90 \pm 0.13$ & ($-$145.15, $-$144.66) & uniform \\
Barycenter $\mu_{\delta}$ (mas/yr) & $-135.21 \pm 0.32$ & ($-$135.8, $-$134.55) & uniform \\
Radial velocity zero point (m/s) & $-129^{+31}_{-68}$ & ($-$246, 180) & uniform \\
\hline
\multicolumn{4}{c}{Derived Parameters} \\ \hline
Period (yrs)     &    $91 \pm 27$ & (55, 160) \\
Argument of periastron $\omega$ ($^\circ$)\tablenotemark{\rm *}   &      ${151}_{-131}^{+155}$ & (4, 356) \\
Eccentricity $e$    &    ${0.18}_{-0.14}^{+0.20}$ & $<$0.46 \\
Semimajor axis (mas)  &       $800^{+150}_{-170}$ & (570, 1160) \\
Periastron time $T_0$ (JD)    &     ${2440900}_{-13100}^{+4500}$ & (2404000, 2453500) \\
Mass ratio   &      ${0.0402}_{-0.0073}^{+0.0078}$ & (0.0267, 0.0563)
\enddata
\tablenotetext{*}{Orbital parameters listed are of HD 33632 Ab.  HD 33632 Aa has $\omega$ shifted by 180$^\circ$. }
\label{tab:mcmc_results}
\end{deluxetable*}

Figure \ref{fig:orbits} shows our posterior distributions for selected orbital parameters, the primary mass (almost identical to our prior of $1.1\pm0.1~M_\odot$), and HD 33632 Ab's mass.  Table \ref{tab:mcmc_results} lists all priors and posteriors from the fit.
The companion has a best-fit semimajor axis of 21.2$^{+4.1}_{-4.5}$ au with an orbit inclined by $i$ = 39$.\!\!^\circ$4$^{+8.0}_{-20}$.   The semimajor axis is double-peaked because our relative astrometry spans only a 2 year baseline, while 25 years elapse between absolute astrometric measurements: {\it Hipparcos} mainly constrains HD 33632 Ab's 1991 position.  Continued orbital monitoring will remove the double peak, improving the companion's mass measurement.  HD 33632 Ab's median posterior eccentricity is $e$ = 0.19 but the distribution peaks at zero (i.e.~a circular orbit): the 68\% and 95\% upper limits are 0.29 and 0.46, respectively.

HD 33632 Ab has a dynamical mass of 46.4$^{+8.1}_{-7.6}$ $M_{\rm J}$ and a mass ratio of $q$ $\sim$ 4.0\% $\pm$ 0.7\%.   We may compare this to masses derived by matching its observed luminosity and age with substellar evolutionary models.  For its full (preferred) age range of 1.5$^{+3.0}_{-0.7}$ Gyr (1--2.5 Gyr),
%
the \citet{Baraffe2003} and \citet{Burrows1997} luminosity evolution models imply masses of 48$^{+17}_{-9}$ $M_{\rm J}$ (41--60 $M_{\rm J}$) and 55$^{+18}_{-15}$ $M_{\rm J}$ (45--65 $M_{\rm J}$), respectively.  
These model-dependent masses agree well with the dynamical mass if we adopt our preferred age of 1--2.5~Gyr, but show moderate tension for ages $\gtrsim$2.5~Gyr.



\section{Discussion}
Our SCExAO/CHARIS and complementary Keck/NIRC2 imaging identify a low-mass ratio brown dwarf companion at $r_{\rm proj}$ $\sim$20 au around the main sequence Sun-like star, HD 33632 Aa.   HD 33632 Ab's infrared colors and spectrum are best matched by a field L/T transition object.   Combining measurements from {\it Hipparcos}, {\it Gaia}, and precision radial-velocity with our CHARIS/NIRC2 relative astrometry yields a dynamical mass of 46.4$^{+8.1}_{-7.6}$ $M_{\rm J}$.  This is more precise than estimates from luminosity evolution models alone, given age uncertainties, and it will improve with continued astrometric monitoring and future {\it Gaia} data releases.  Provided that the system age and companion mass are better constrained in future work, HD~33632~Ab adds to the still-small sample of objects enabling us to test substellar evolutionary models across masses and ages.

HD 33632 Ab lies at the field L/T transition tracking the dissipation of clouds/dust in substellar atmospheres.   In $L_{\rm p}$/$H$-$L_{\rm p}$, its colors overlap with and its temperature is just slightly exceeds those of the young exoplanets HR 8799 cde.    HD 33632 Ab represents an older, higher mass, higher gravity, and (likely) less cloudy/dusty counterpart to some of the first imaged exoplanets.   Thermal infrared follow-up imaging and spectroscopy on the ground and with the \textit{James Webb Space Telescope} could further clarify how its atmospheric chemistry compares with HR 8799 cde and other companions.   Theoretical modeling incorporating these higher-resolution data will quantify constraints on HD 33632 Ab's atmosphere.

HD 33632 Ab's mass, mass ratio, and orbital separation are nearly identical to those of the older, colder brown dwarf companion GJ 758 B \citep{Thalmann2009,Brandt2019}, which likewise orbits a Sun-like star.   Its mass and mass ratio bracket those of other companions to mid F to early G stars orbiting at 20--50 au: e.g.~the more massive, older T dwarf companion HD 19467~B
\citep[][]{Crepp2014} and (near-) planet mass companions HR 2562~b \citep[]{Konopacky2016} and GJ 504~b \citep[]{Kuzuhara2013} with comparable or younger ages.  

HD~33632~Ab's eccentricity posterior probability distribution has a 68\% upper limit of 0.29.  This is smaller than the eccentricities of most brown dwarfs studied in \citet{Bowler2020}, who argue that systematically higher eccentricities of brown dwarf-mass companions relative to exoplanets point to a different formation mechanism.    Future CHARIS/NIRC2 astrometry for HD 33632 Ab together with more precise HD 33632 Aa astrometry in future {\it Gaia} data releases will better constrain the companion's mass and eccentricity. 

In addition to HD 33632 Aa, hundreds of nearby stars show evidence for a statistically significant acceleration plausibly caused by an unidentified substellar companion.   Upcoming {\it Gaia} data releases will identify thousands more accelerating stars.  Some will be young enough that extreme AO systems like SCExAO can image self-luminous jovian exoplanets responsible for the astrometric trends.   Our work is a proof-in-concept that direct imaging searches targeting HGCA-selected stars may 
identify new substellar and potentially even planetary companions, increasing the yield of discoveries.

 
\acknowledgments
\indent We thank Eric Mamajek and William Cochran for helpful comments regarding the HD 33632 system properties.
\indent The authors acknowledge the very significant cultural role and reverence that the summit of Mauna Kea holds within the Hawaiian community.  We are most fortunate to have the opportunity to conduct observations from this mountain.   

\indent We acknowledge the critical importance of the current and recent Subaru and Keck Observatory daycrew, technicians, support astronomers, telescope operators, computer support, and office staff employees, especially during the challenging times presented by the COVID-19 pandemic.  Their expertise, ingenuity, and dedication is indispensable to the continued successful operation of these observatories.  
\\
\indent We thank the Subaru and NASA Keck Time Allocation Committees for their generous support of this program.  TC was supported by a NASA Senior Postdoctoral Fellowship and NASA/Keck grant LK-2663-948181.   TB gratefully
acknowledges support from the Heising-Simons foundation and from NASA under grant \#80NSSC18K0439.  MT is supported by JSPS KAKENHI Grant \# 18H05442.
\\
\indent The development of SCExAO was supported by JSPS (Grant-in-Aid for Research \#23340051, \#26220704 \& \#23103002), Astrobiology Center of NINS, Japan, the Mt Cuba Foundation, and the director's contingency fund at Subaru Telescope.  CHARIS was developed under the support by the Grant-in-Aid for Scientific Research on Innovative Areas \#2302.  Some of the data presented herein were obtained at the W. M. Keck Observatory, which is operated as a scientific partnership among the California Institute of Technology, the University of California and the National Aeronautics and Space Administration. The Observatory was made possible by the generous financial support of the W. M. Keck Foundation.\\


\begin{thebibliography}{}
\expandafter\ifx\csname natexlab\endcsname\relax\def\natexlab#1{#1}\fi
\providecommand{\url}[1]{\href{#1}{#1}}
\providecommand{\dodoi}[1]{doi:~\href{http://doi.org/#1}{\nolinkurl{#1}}}
\providecommand{\doeprint}[1]{\href{http://ascl.net/#1}{\nolinkurl{http://ascl.net/#1}}}
\providecommand{\doarXiv}[1]{\href{https://arxiv.org/abs/#1}{\nolinkurl{https://arxiv.org/abs/#1}}}

\bibitem[{{Baraffe} {et~al.}(2003){Baraffe}, {Chabrier}, {Barman}, {Allard}, \&
  {Hauschildt}}]{Baraffe2003}
{Baraffe}, I., {Chabrier}, G., {Barman}, T.~S., {Allard}, F., \& {Hauschildt},
  P.~H. 2003, \aap, 402, 701, \dodoi{10.1051/0004-6361:20030252}

\bibitem[{{Barman} {et~al.}(2015){Barman}, {Konopacky}, {Macintosh}, \&
  {Marois}}]{Barman2015}
{Barman}, T.~S., {Konopacky}, Q.~M., {Macintosh}, B., \& {Marois}, C. 2015,
  \apj, 804, 61, \dodoi{10.1088/0004-637X/804/1/61}

\bibitem[{{Bowler} {et~al.}(2020){Bowler}, {Blunt}, \& {Nielsen}}]{Bowler2020}
{Bowler}, B.~P., {Blunt}, S.~C., \& {Nielsen}, E.~L. 2020, \aj, 159, 63,
  \dodoi{10.3847/1538-3881/ab5b11}

\bibitem[{{Brandt}(2018)}]{Brandt2018}
{Brandt}, T.~D. 2018, \apjs, 239, 31, \dodoi{10.3847/1538-4365/aaec06}

\bibitem[{{Brandt} {et~al.}(2019){Brandt}, {Dupuy}, \& {Bowler}}]{Brandt2019}
{Brandt}, T.~D., {Dupuy}, T.~J., \& {Bowler}, B.~P. 2019, \aj, 158, 140,
  \dodoi{10.3847/1538-3881/ab04a8}

\bibitem[{{Brandt} \& {Huang}(2015)}]{Brandt2015}
{Brandt}, T.~D., \& {Huang}, C.~X. 2015, \apj, 807, 24,
  \dodoi{10.1088/0004-637X/807/1/24}

\bibitem[{{Brandt} {et~al.}(2014){Brandt}, {McElwain}, {Turner}, {Mede},
  {Spiegel}, {Kuzuhara}, {Schlieder}, {Wisniewski}, {Abe}, {Biller},
  {Brandner}, {Carson}, {Currie}, {Egner}, {Feldt}, {Golota}, {Goto}, {Grady},
  {Guyon}, {Hashimoto}, {Hayano}, {Hayashi}, {Hayashi}, {Henning}, {Hodapp},
  {Inutsuka}, {Ishii}, {Iye}, {Janson}, {Kand ori}, {Knapp}, {Kudo},
  {Kusakabe}, {Kwon}, {Matsuo}, {Miyama}, {Morino}, {Moro-Mart{\'\i}n},
  {Nishimura}, {Pyo}, {Serabyn}, {Suto}, {Suzuki}, {Takami}, {Takato},
  {Terada}, {Thalmann}, {Tomono}, {Watanabe}, {Yamada}, {Takami}, {Usuda}, \&
  {Tamura}}]{Brandt2014}
{Brandt}, T.~D., {McElwain}, M.~W., {Turner}, E.~L., {et~al.} 2014, \apj, 794,
  159, \dodoi{10.1088/0004-637X/794/2/159}

\bibitem[{{Brandt} {et~al.}(2017){Brandt}, {Rizzo}, {Groff}, {Chilcote},
  {Greco}, {Kasdin}, {Limbach}, {Galvin}, {Loomis}, {Knapp}, {McElwain},
  {Jovanovic}, {Currie}, {Mede}, {Tamura}, {Takato}, \& {Hayashi}}]{Brandt2017}
{Brandt}, T.~D., {Rizzo}, M., {Groff}, T., {et~al.} 2017, Journal of
  Astronomical Telescopes, Instruments, and Systems, 3, 048002,
  \dodoi{10.1117/1.JATIS.3.4.048002}

\bibitem[{{Burrows} {et~al.}(2006){Burrows}, {Sudarsky}, \&
  {Hubeny}}]{Burrows2006}
{Burrows}, A., {Sudarsky}, D., \& {Hubeny}, I. 2006, \apj, 640, 1063,
  \dodoi{10.1086/500293}

\bibitem[{{Burrows} {et~al.}(1997){Burrows}, {Marley}, {Hubbard}, {Lunine},
  {Guillot}, {Saumon}, {Freedman}, {Sudarsky}, \& {Sharp}}]{Burrows1997}
{Burrows}, A., {Marley}, M., {Hubbard}, W.~B., {et~al.} 1997, \apj, 491, 856,
  \dodoi{10.1086/305002}


\bibitem[{{Castro} {et~al.}(2016){Castro}, {Duarte}, {Pace}, \& {do
  Nascimento}}]{Castro2016}
{Castro}, M., {Duarte}, T., {Pace}, G., \& {do Nascimento}, J.~D. 2016, \aap,
  590, A94, \dodoi{10.1051/0004-6361/201527583}

\bibitem[{{Crepp} \& {Johnson}(2011)}]{Crepp2011}
{Crepp}, J.~R., \& {Johnson}, J.~A. 2011, \apj, 733, 126,
  \dodoi{10.1088/0004-637X/733/2/126}

\bibitem[{{Crepp} {et~al.}(2014){Crepp}, {Johnson}, {Howard}, {Marcy},
  {Brewer}, {Fischer}, {Wright}, \& {Isaacson}}]{Crepp2014}
{Crepp}, J.~R., {Johnson}, J.~A., {Howard}, A.~W., {et~al.} 2014, \apj, 781,
  29, \dodoi{10.1088/0004-637X/781/1/29}

\bibitem[{{Currie} {et~al.}(2015){Currie}, {Cloutier}, {Brittain}, {Grady},
  {Burrows}, {Muto}, {Kenyon}, \& {Kuchner}}]{Currie2015}
{Currie}, T., {Cloutier}, R., {Brittain}, S., {et~al.} 2015, \apjl, 814, L27,
  \dodoi{10.1088/2041-8205/814/2/L27}

\bibitem[{{Currie} {et~al.}(2014{\natexlab{a}}){Currie}, {Daemgen}, {Debes},
  {Lafreniere}, {Itoh}, {Jayawardhana}, {Ratzka}, \& {Correia}}]{Currie2014a}
{Currie}, T., {Daemgen}, S., {Debes}, J., {et~al.} 2014{\natexlab{a}}, \apjl,
  780, L30, \dodoi{10.1088/2041-8205/780/2/L30}

\bibitem[{{Currie} {et~al.}(2011){Currie}, {Burrows}, {Itoh}, {Matsumura},
  {Fukagawa}, {Apai}, {Madhusudhan}, {Hinz}, {Rodigas}, {Kasper}, {Pyo}, \&
  {Ogino}}]{Currie2011}
{Currie}, T., {Burrows}, A., {Itoh}, Y., {et~al.} 2011, \apj, 729, 128,
  \dodoi{10.1088/0004-637X/729/2/128}

\bibitem[{{Currie} {et~al.}(2014{\natexlab{b}}){Currie}, {Burrows}, {Girard},
  {Cloutier}, {Fukagawa}, {Sorahana}, {Kuchner}, {Kenyon}, {Madhusudhan},
  {Itoh}, {Jayawardhana}, {Matsumura}, \& {Pyo}}]{Currie2014b}
{Currie}, T., {Burrows}, A., {Girard}, J.~H., {et~al.} 2014{\natexlab{b}},
  \apj, 795, 133, \dodoi{10.1088/0004-637X/795/2/133}

\bibitem[{{Currie} {et~al.}(2018){Currie}, {Brandt}, {Uyama}, {Nielsen},
  {Blunt}, {Guyon}, {Tamura}, {Marois}, {Mede}, {Kuzuhara}, {Groff},
  {Jovanovic}, {Kasdin}, {Lozi}, {Hodapp}, {Chilcote}, {Carson}, {Martinache},
  {Goebel}, {Grady}, {McElwain}, {Akiyama}, {Asensio-Torres}, {Hayashi},
  {Janson}, {Knapp}, {Kwon}, {Nishikawa}, {Oh}, {Schlieder}, {Serabyn},
  {Sitko}, \& {Skaf}}]{Currie2018}
{Currie}, T., {Brandt}, T.~D., {Uyama}, T., {et~al.} 2018, \aj, 156, 291,
  \dodoi{10.3847/1538-3881/aae9ea}

\bibitem[{{Dupuy} {et~al.}(2019){Dupuy}, {Brandt}, {Kratter}, \&
  {Bowler}}]{Dupuy2019}
{Dupuy}, T.~J., {Brandt}, T.~D., {Kratter}, K.~M., \& {Bowler}, B.~P. 2019,
  \apjl, 871, L4, \dodoi{10.3847/2041-8213/aafb31}
  
\bibitem[{{Dupuy} \& {Liu}(2012)}]{DupuyLiu2012}
{Dupuy}, T.~J., \& {Liu}, M.~C. 2012, \apjs, 201, 19,
  \dodoi{10.1088/0067-0049/201/2/19}
  
  

\bibitem[{{Fernandes} {et~al.}(2019){Fernandes}, {Mulders}, {Pascucci},
  {Mordasini}, \& {Emsenhuber}}]{Fernandes2019}
{Fernandes}, R.~B., {Mulders}, G.~D., {Pascucci}, I., {Mordasini}, C., \&
  {Emsenhuber}, A. 2019, \apj, 874, 81, \dodoi{10.3847/1538-4357/ab0300}

\bibitem[{{Fischer} {et~al.}(2014){Fischer}, {Marcy}, \&
  {Spronck}}]{Fischer2014}
{Fischer}, D.~A., {Marcy}, G.~W., \& {Spronck}, J. F.~P. 2014, \apjs, 210, 5,
  \dodoi{10.1088/0067-0049/210/1/5}

\bibitem[{{Freund} {et~al.}(2020){Freund}, {Robrade}, {Schneider}, \&
  {Schmitt}}]{Freund2020}
{Freund}, S., {Robrade}, J., {Schneider}, P.~C., \& {Schmitt}, J.~H.~M.~M.
  2020, \aap, 640, A66, \dodoi{10.1051/0004-6361/201937304}

\bibitem[{{Gagn{\'e}} {et~al.}(2014){Gagn{\'e}}, {Lafreni{\`e}re}, {Doyon},
  {Malo}, \& {Artigau}}]{Gagne2014}
{Gagn{\'e}}, J., {Lafreni{\`e}re}, D., {Doyon}, R., {Malo}, L., \& {Artigau},
  {\'E}. 2014, \apj, 783, 121, \dodoi{10.1088/0004-637X/783/2/121}

\bibitem[{{Gaia Collaboration} {et~al.}(2018){Gaia Collaboration}, {Brown},
  {Vallenari}, {Prusti}, {de Bruijne}, {Babusiaux}, {Bailer-Jones}, {Biermann},
  {Evans}, {Eyer}, {Jansen}, {Jordi}, {Klioner}, {Lammers}, {Lindegren},
  {Luri}, {Mignard}, {Panem}, {Pourbaix}, {Randich}, {Sartoretti}, {Siddiqui},
  {Soubiran}, {van Leeuwen}, {Walton}, {Arenou}, {Bastian}, {Cropper},
  {Drimmel}, {Katz}, {Lattanzi}, {Bakker}, {Cacciari}, {Casta{\~n}eda},
  {Chaoul}, {Cheek}, {De Angeli}, {Fabricius}, {Guerra}, {Holl}, {Masana},
  {Messineo}, {Mowlavi}, {Nienartowicz}, {Panuzzo}, {Portell}, {Riello},
  {Seabroke}, {Tanga}, {Th{\'e}venin}, {Gracia-Abril}, {Comoretto},
  {Garcia-Reinaldos}, {Teyssier}, {Altmann}, {Andrae}, {Audard},
  {Bellas-Velidis}, {Benson}, {Berthier}, {Blomme}, {Burgess}, {Busso},
  {Carry}, {Cellino}, {Clementini}, {Clotet}, {Creevey}, {Davidson}, {De
  Ridder}, {Delchambre}, {Dell'Oro}, {Ducourant},
  {Fern{\'a}ndez-Hern{\'a}ndez}, {Fouesneau}, {Fr{\'e}mat}, {Galluccio},
  {Garc{\'\i}a-Torres}, {Gonz{\'a}lez-N{\'u}{\~n}ez}, {Gonz{\'a}lez-Vidal},
  {Gosset}, {Guy}, {Halbwachs}, {Hambly}, {Harrison}, {Hern{\'a}ndez},
  {Hestroffer}, {Hodgkin}, {Hutton}, {Jasniewicz}, {Jean-Antoine-Piccolo},
  {Jordan}, {Korn}, {Krone-Martins}, {Lanzafame}, {Lebzelter}, {L{\"o}ffler},
  {Manteiga}, {Marrese}, {Mart{\'\i}n-Fleitas}, {Moitinho}, {Mora}, {Muinonen},
  {Osinde}, {Pancino}, {Pauwels}, {Petit}, {Recio-Blanco}, {Richards},
  {Rimoldini}, {Robin}, {Sarro}, {Siopis}, {Smith}, {Sozzetti}, {S{\"u}veges},
  {Torra}, {van Reeven}, {Abbas}, {Abreu Aramburu}, {Accart}, {Aerts},
  {Altavilla}, {{\'A}lvarez}, {Alvarez}, {Alves}, {Anderson}, {Andrei},
  {Anglada Varela}, {Antiche}, {Antoja}, {Arcay}, {Astraatmadja}, {Bach},
  {Baker}, {Balaguer-N{\'u}{\~n}ez}, {Balm}, {Barache}, {Barata}, {Barbato},
  {Barblan}, {Barklem}, {Barrado}, {Barros}, {Barstow}, {Bartholom{\'e}
  Mu{\~n}oz}, {Bassilana}, {Becciani}, {Bellazzini}, {Berihuete}, {Bertone},
  {Bianchi}, {Bienaym{\'e}}, {Blanco-Cuaresma}, {Boch}, {Boeche}, {Bombrun},
  {Borrachero}, {Bossini}, {Bouquillon}, {Bourda}, {Bragaglia}, {Bramante},
  {Breddels}, {Bressan}, {Brouillet}, {Br{\"u}semeister}, {Brugaletta},
  {Bucciarelli}, {Burlacu}, {Busonero}, {Butkevich}, {Buzzi}, {Caffau},
  {Cancelliere}, {Cannizzaro}, {Cantat-Gaudin}, {Carballo}, {Carlucci},
  {Carrasco}, {Casamiquela}, {Castellani}, {Castro-Ginard}, {Charlot},
  {Chemin}, {Chiavassa}, {Cocozza}, {Costigan}, {Cowell}, {Crifo}, {Crosta},
  {Crowley}, {Cuypers}, {Dafonte}, {Damerdji}, {Dapergolas}, {David}, {David},
  {de Laverny}, {De Luise}, {De March}, {de Martino}, {de Souza}, {de Torres},
  {Debosscher}, {del Pozo}, {Delbo}, {Delgado}, {Delgado}, {Di Matteo},
  {Diakite}, {Diener}, {Distefano}, {Dolding}, {Drazinos}, {Dur{\'a}n},
  {Edvardsson}, {Enke}, {Eriksson}, {Esquej}, {Eynard Bontemps}, {Fabre},
  {Fabrizio}, {Faigler}, {Falc{\~a}o}, {Farr{\`a}s Casas}, {Federici},
  {Fedorets}, {Fernique}, {Figueras}, {Filippi}, {Findeisen}, {Fonti},
  {Fraile}, {Fraser}, {Fr{\'e}zouls}, {Gai}, {Galleti}, {Garabato},
  {Garc{\'\i}a-Sedano}, {Garofalo}, {Garralda}, {Gavel}, {Gavras}, {Gerssen},
  {Geyer}, {Giacobbe}, {Gilmore}, {Girona}, {Giuffrida}, {Glass}, {Gomes},
  {Granvik}, {Gueguen}, {Guerrier}, {Guiraud}, {Guti{\'e}rrez-S{\'a}nchez},
  {Haigron}, {Hatzidimitriou}, {Hauser}, {Haywood}, {Heiter}, {Helmi}, {Heu},
  {Hilger}, {Hobbs}, {Hofmann}, {Holland}, {Huckle}, {Hypki}, {Icardi},
  {Jan{\ss}en}, {Jevardat de Fombelle}, {Jonker}, {Juh{\'a}sz}, {Julbe},
  {Karampelas}, {Kewley}, {Klar}, {Kochoska}, {Kohley}, {Kolenberg},
  {Kontizas}, {Kontizas}, {Koposov}, {Kordopatis}, {Kostrzewa-Rutkowska},
  {Koubsky}, {Lambert}, {Lanza}, {Lasne}, {Lavigne}, {Le Fustec}, {Le
  Poncin-Lafitte}, {Lebreton}, {Leccia}, {Leclerc}, {Lecoeur-Taibi},
  {Lenhardt}, {Leroux}, {Liao}, {Licata}, {Lindstr{\o}m}, {Lister}, {Livanou},
  {Lobel}, {L{\'o}pez}, {Managau}, {Mann}, {Mantelet}, {Marchal}, {Marchant},
  {Marconi}, {Marinoni}, {Marschalk{\'o}}, {Marshall}, {Martino}, {Marton},
  {Mary}, {Massari}, {Matijevi{\v{c}}}, {Mazeh}, {McMillan}, {Messina},
  {Michalik}, {Millar}, {Molina}, {Molinaro}, {Moln{\'a}r}, {Montegriffo},
  {Mor}, {Morbidelli}, {Morel}, {Morris}, {Mulone}, {Muraveva}, {Musella},
  {Nelemans}, {Nicastro}, {Noval}, {O'Mullane}, {Ord{\'e}novic},
  {Ord{\'o}{\~n}ez-Blanco}, {Osborne}, {Pagani}, {Pagano}, {Pailler},
  {Palacin}, {Palaversa}, {Panahi}, {Pawlak}, {Piersimoni}, {Pineau}, {Plachy},
  {Plum}, {Poggio}, {Poujoulet}, {Pr{\v{s}}a}, {Pulone}, {Racero}, {Ragaini},
  {Rambaux}, {Ramos-Lerate}, {Regibo}, {Reyl{\'e}}, {Riclet}, {Ripepi}, {Riva},
  {Rivard}, {Rixon}, {Roegiers}, {Roelens}, {Romero-G{\'o}mez}, {Rowell},
  {Royer}, {Ruiz-Dern}, {Sadowski}, {Sagrist{\`a} Sell{\'e}s}, {Sahlmann},
  {Salgado}, {Salguero}, {Sanna}, {Santana-Ros}, {Sarasso}, {Savietto},
  {Schultheis}, {Sciacca}, {Segol}, {Segovia}, {S{\'e}gransan}, {Shih},
  {Siltala}, {Silva}, {Smart}, {Smith}, {Solano}, {Solitro}, {Sordo}, {Soria
  Nieto}, {Souchay}, {Spagna}, {Spoto}, {Stampa}, {Steele},
  {Steidelm{\"u}ller}, {Stephenson}, {Stoev}, {Suess}, {Surdej}, {Szabados},
  {Szegedi-Elek}, {Tapiador}, {Taris}, {Tauran}, {Taylor}, {Teixeira},
  {Terrett}, {Teyssand ier}, {Thuillot}, {Titarenko}, {Torra Clotet}, {Turon},
  {Ulla}, {Utrilla}, {Uzzi}, {Vaillant}, {Valentini}, {Valette}, {van Elteren},
  {Van Hemelryck}, {van Leeuwen}, {Vaschetto}, {Vecchiato}, {Veljanoski},
  {Viala}, {Vicente}, {Vogt}, {von Essen}, {Voss}, {Votruba}, {Voutsinas},
  {Walmsley}, {Weiler}, {Wertz}, {Wevers}, {Wyrzykowski}, {Yoldas},
  {{\v{Z}}erjal}, {Ziaeepour}, {Zorec}, {Zschocke}, {Zucker}, {Zurbach}, \&
  {Zwitter}}]{GAIA2018}
{Gaia Collaboration}, {Brown}, A.~G.~A., {Vallenari}, A., {et~al.} 2018, \aap,
  616, A1, \dodoi{10.1051/0004-6361/201833051}

\bibitem[{{Golimowski} {et~al.}(2004){Golimowski}, {Leggett}, {Marley}, {Fan},
  {Geballe}, {Knapp}, {Vrba}, {Henden}, {Luginbuhl}, {Guetter}, {Munn},
  {Canzian}, {Zheng}, {Tsvetanov}, {Chiu}, {Glazebrook}, {Hoversten},
  {Schneider}, \& {Brinkmann}}]{Golimowski2004}
{Golimowski}, D.~A., {Leggett}, S.~K., {Marley}, M.~S., {et~al.} 2004, \aj,
  127, 3516, \dodoi{10.1086/420709}

\bibitem[{{Greco} \& {Brandt}(2016)}]{GrecoBrandt2016}
{Greco}, J.~P., \& {Brandt}, T.~D. 2016, \apj, 833, 134,
  \dodoi{10.3847/1538-4357/833/2/134}

\bibitem[{{Groff} {et~al.}(2016){Groff}, {Chilcote}, {Kasdin}, {Galvin},
  {Loomis}, {Carr}, {Brand t}, {Knapp}, {Limbach}, {Guyon}, {Jovanovic},
  {McElwain}, {Takato}, \& {Hayashi}}]{Groff2016}
{Groff}, T.~D., {Chilcote}, J., {Kasdin}, N.~J., {et~al.} 2016, in Society of
  Photo-Optical Instrumentation Engineers (SPIE) Conference Series, Vol. 9908,
  Ground-based and Airborne Instrumentation for Astronomy VI, 99080O,
  \dodoi{10.1117/12.2233447}

\bibitem[{{Jovanovic} {et~al.}(2015{\natexlab{a}}){Jovanovic}, {Guyon},
  {Martinache}, {Pathak}, {Hagelberg}, \& {Kudo}}]{Jovanovic2015-astrogrids}
{Jovanovic}, N., {Guyon}, O., {Martinache}, F., {et~al.} 2015{\natexlab{a}},
  \apjl, 813, L24, \dodoi{10.1088/2041-8205/813/2/L24}

\bibitem[{{Jovanovic} {et~al.}(2015{\natexlab{b}}){Jovanovic}, {Martinache},
  {Guyon}, {Clergeon}, {Singh}, {Kudo}, {Garrel}, {Newman}, {Doughty}, {Lozi},
  {Males}, {Minowa}, {Hayano}, {Takato}, {Morino}, {Kuhn}, {Serabyn}, {Norris},
  {Tuthill}, {Schworer}, {Stewart}, {Close}, {Huby}, {Perrin}, {Lacour},
  {Gauchet}, {Vievard}, {Murakami}, {Oshiyama}, {Baba}, {Matsuo}, {Nishikawa},
  {Tamura}, {Lai}, {Marchis}, {Duchene}, {Kotani}, \&
  {Woillez}}]{Jovanovic2015}
{Jovanovic}, N., {Martinache}, F., {Guyon}, O., {et~al.} 2015{\natexlab{b}},
  \pasp, 127, 890, \dodoi{10.1086/682989}

\bibitem[{{Keppler} {et~al.}(2018){Keppler}, {Benisty}, {M{\"u}ller},
  {Henning}, {van Boekel}, {Cantalloube}, {Ginski}, {van Holstein}, {Maire},
  {Pohl}, {Samland }, {Avenhaus}, {Baudino}, {Boccaletti}, {de Boer},
 {Bonnefoy}, {Chauvin}, {Desidera}, {Langlois}, {Lazzoni}, {Marleau},
  {Mordasini}, {Pawellek}, {Stolker}, {Vigan}, {Zurlo}, {Birnstiel},
  {Brandner}, {Feldt}, {Flock}, {Girard}, {Gratton}, {Hagelberg}, {Isella},
  {Janson}, {Juhasz}, {Kemmer}, {Kral}, {Lagrange}, {Launhardt}, {Matter},
  {M{\'e}nard}, {Milli}, {Molli{\`e}re}, {Olofsson}, {P{\'e}rez}, {Pinilla},
  {Pinte}, {Quanz}, {Schmidt}, {Udry}, {Wahhaj}, {Williams}, {Buenzli},
  {Cudel}, {Dominik}, {Galicher}, {Kasper}, {Lannier}, {Mesa}, {Mouillet},
  {Peretti}, {Perrot}, {Salter}, {Sissa}, {Wildi}, {Abe}, {Antichi},
  {Augereau}, {Baruffolo}, {Baudoz}, {Bazzon}, {Beuzit}, {Blanchard}, {Brems},
  {Buey}, {De Caprio}, {Carbillet}, {Carle}, {Cascone}, {Cheetham}, {Claudi},
  {Costille}, {Delboulb{\'e}}, {Dohlen}, {Fantinel}, {Feautrier}, {Fusco},
  {Giro}, {Gluck}, {Gry}, {Hubin}, {Hugot}, {Jaquet}, {Le Mignant}, {Llored},
  {Madec}, {Magnard}, {Martinez}, {Maurel}, {Meyer}, {M{\"o}ller-Nilsson},
  {Moulin}, {Mugnier}, {Orign{\'e}}, {Pavlov}, {Perret}, {Petit}, {Pragt},
  {Puget}, {Rabou}, {Ramos}, {Rigal}, {Rochat}, {Roelfsema}, {Rousset}, {Roux},
  {Salasnich}, {Sauvage}, {Sevin}, {Soenke}, {Stadler}, {Suarez}, {Turatto}, \&
  {Weber}}]{Keppler2018}
{Keppler}, M., {Benisty}, M., {M{\"u}ller}, A., {et~al.} 2018, \aap, 617, A44,
  \dodoi{10.1051/0004-6361/201832957}

\bibitem[{{Konopacky} {et~al.}(2016){Konopacky}, {Rameau}, {Duch{\^e}ne},
  {Filippazzo}, {Giorla Godfrey}, {Marois}, {Nielsen}, {Pueyo}, {Rafikov},
  {Rice}, {Wang}, {Ammons}, {Bailey}, {Barman}, {Bulger}, {Bruzzone},
  {Chilcote}, {Cotten}, {Dawson}, {De Rosa}, {Doyon}, {Esposito}, {Fitzgerald},
  {Follette}, {Goodsell}, {Graham}, {Greenbaum}, {Hibon}, {Hung}, {Ingraham},
  {Kalas}, {Lafreni{\`e}re}, {Larkin}, {Macintosh}, {Maire}, {Marchis},
  {Marley}, {Matthews}, {Metchev}, {Millar-Blanchaer}, {Oppenheimer}, {Palmer},
  {Patience}, {Perrin}, {Poyneer}, {Rajan}, {Rantakyr{\"o}}, {Savransky},
  {Schneider}, {Sivaramakrishnan}, {Song}, {Soummer}, {Thomas}, {Wallace},
  {Ward-Duong}, {Wiktorowicz}, \& {Wolff}}]{Konopacky2016}
{Konopacky}, Q.~M., {Rameau}, J., {Duch{\^e}ne}, G., {et~al.} 2016, \apjl, 829,
  L4, \dodoi{10.3847/2041-8205/829/1/L4}

\bibitem[{{Kuzuhara} {et~al.}(2013){Kuzuhara}, {Tamura}, {Kudo}, {Janson},
  {Kand ori}, {Brandt}, {Thalmann}, {Spiegel}, {Biller}, {Carson}, {Hori},
  {Suzuki}, {Burrows}, {Henning}, {Turner}, {McElwain}, {Moro-Mart{\'\i}n},
  {Suenaga}, {Takahashi}, {Kwon}, {Lucas}, {Abe}, {Brand ner}, {Egner},
  {Feldt}, {Fujiwara}, {Goto}, {Grady}, {Guyon}, {Hashimoto}, {Hayano},
  {Hayashi}, {Hayashi}, {Hodapp}, {Ishii}, {Iye}, {Knapp}, {Matsuo}, {Mayama},
  {Miyama}, {Morino}, {Nishikawa}, {Nishimura}, {Kotani}, {Kusakabe}, {Pyo},
  {Serabyn}, {Suto}, {Takami}, {Takato}, {Terada}, {Tomono}, {Watanabe},
  {Wisniewski}, {Yamada}, {Takami}, \& {Usuda}}]{Kuzuhara2013}
{Kuzuhara}, M., {Tamura}, M., {Kudo}, T., {et~al.} 2013, \apj, 774, 11,
  \dodoi{10.1088/0004-637X/774/1/11}

\bibitem[{{Macintosh} {et~al.}(2015){Macintosh}, {Graham}, {Barman}, {De Rosa},
  {Konopacky}, {Marley}, {Marois}, {Nielsen}, {Pueyo}, {Rajan}, {Rameau},
  {Saumon}, {Wang}, {Patience}, {Ammons}, {Arriaga}, {Artigau}, {Beckwith},
  {Brewster}, {Bruzzone}, {Bulger}, {Burningham}, {Burrows}, {Chen}, {Chiang},
  {Chilcote}, {Dawson}, {Dong}, {Doyon}, {Draper}, {Duch{\^e}ne}, {Esposito},
  {Fabrycky}, {Fitzgerald}, {Follette}, {Fortney}, {Gerard}, {Goodsell},
  {Greenbaum}, {Hibon}, {Hinkley}, {Cotten}, {Hung}, {Ingraham},
  {Johnson-Groh}, {Kalas}, {Lafreniere}, {Larkin}, {Lee}, {Line}, {Long},
  {Maire}, {Marchis}, {Matthews}, {Max}, {Metchev}, {Millar-Blanchaer},
  {Mittal}, {Morley}, {Morzinski}, {Murray-Clay}, {Oppenheimer}, {Palmer},
  {Patel}, {Perrin}, {Poyneer}, {Rafikov}, {Rantakyr{\"o}}, {Rice}, {Rojo},
  {Rudy}, {Ruffio}, {Ruiz}, {Sadakuni}, {Saddlemyer}, {Salama}, {Savransky},
  {Schneider}, {Sivaramakrishnan}, {Song}, {Soummer}, {Thomas}, {Vasisht},
  {Wallace}, {Ward-Duong}, {Wiktorowicz}, {Wolff}, \&
  {Zuckerman}}]{Macintosh2015}
{Macintosh}, B., {Graham}, J.~R., {Barman}, T., {et~al.} 2015, Science, 350,
  64, \dodoi{10.1126/science.aac5891}

\bibitem[{{Mamajek} \& {Hillenbrand}(2008)}]{Mamajek2008}
{Mamajek}, E.~E., \& {Hillenbrand}, L.~A. 2008, \apj, 687, 1264,
  \dodoi{10.1086/591785}

\bibitem[{{Marois} {et~al.}(2000){Marois}, {Doyon}, {Racine}, \&
  {Nadeau}}]{Marois2000}
{Marois}, C., {Doyon}, R., {Racine}, R., \& {Nadeau}, D. 2000, \pasp, 112, 91,
  \dodoi{10.1086/316492}

\bibitem[{{Marois} {et~al.}(2006){Marois}, {Lafreni{\`e}re}, {Doyon},
  {Macintosh}, \& {Nadeau}}]{Marois2006}
{Marois}, C., {Lafreni{\`e}re}, D., {Doyon}, R., {Macintosh}, B., \& {Nadeau},
  D. 2006, \apj, 641, 556, \dodoi{10.1086/500401}

\bibitem[{{Marois} {et~al.}(2008){Marois}, {Macintosh}, {Barman}, {Zuckerman},
  {Song}, {Patience}, {Lafreni{\`e}re}, \& {Doyon}}]{Marois2008a}
{Marois}, C., {Macintosh}, B., {Barman}, T., {et~al.} 2008, Science, 322, 1348,
  \dodoi{10.1126/science.1166585}

\bibitem[{{Marois} {et~al.}(2010){Marois}, {Zuckerman}, {Konopacky},
  {Macintosh}, \& {Barman}}]{Marois2010}
{Marois}, C., {Zuckerman}, B., {Konopacky}, Q.~M., {Macintosh}, B., \&
  {Barman}, T. 2010, \nat, 468, 1080, \dodoi{10.1038/nature09684}

\bibitem[{{Nielsen} {et~al.}(2012){Nielsen}, {Liu}, {Wahhaj}, {Biller},
  {Hayward}, {Boss}, {Bowler}, {Kraus}, {Shkolnik}, {Tecza}, {Chun}, {Clarke},
  {Close}, {Ftaclas}, {Hartung}, {Males}, {Reid}, {Skemer}, {Alencar},
  {Burrows}, {de Gouveia Dal Pino}, {Gregorio-Hetem}, {Kuchner}, {Thatte}, \&
  {Toomey}}]{Nielsen2012}
{Nielsen}, E.~L., {Liu}, M.~C., {Wahhaj}, Z., {et~al.} 2012, \apj, 750, 53,
  \dodoi{10.1088/0004-637X/750/1/53}

\bibitem[{{Nielsen} {et~al.}(2019){Nielsen}, {De Rosa}, {Macintosh}, {Wang},
  {Ruffio}, {Chiang}, {Marley}, {Saumon}, {Savransky}, {Ammons}, {Bailey},
  {Barman}, {Blain}, {Bulger}, {Burrows}, {Chilcote}, {Cotten}, {Czekala},
  {Doyon}, {Duch{\^e}ne}, {Esposito}, {Fabrycky}, {Fitzgerald}, {Follette},
  {Fortney}, {Gerard}, {Goodsell}, {Graham}, {Greenbaum}, {Hibon}, {Hinkley},
  {Hirsch}, {Hom}, {Hung}, {Dawson}, {Ingraham}, {Kalas}, {Konopacky},
  {Larkin}, {Lee}, {Lin}, {Maire}, {Marchis}, {Marois}, {Metchev},
  {Millar-Blanchaer}, {Morzinski}, {Oppenheimer}, {Palmer}, {Patience},
  {Perrin}, {Poyneer}, {Pueyo}, {Rafikov}, {Rajan}, {Rameau}, {Rantakyr{\"o}},
  {Ren}, {Schneider}, {Sivaramakrishnan}, {Song}, {Soummer}, {Tallis},
  {Thomas}, {Ward-Duong}, \& {Wolff}}]{Nielsen2019}
{Nielsen}, E.~L., {De Rosa}, R.~J., {Macintosh}, B., {et~al.} 2019, \aj, 158,
  13, \dodoi{10.3847/1538-3881/ab16e9}

\bibitem[{{Pace}(2013)}]{Pace2013}
{Pace}, G. 2013, \aap, 551, L8, \dodoi{10.1051/0004-6361/201220364}

\bibitem[{{Pueyo}(2016)}]{Pueyo2016}
{Pueyo}, L. 2016, \apj, 824, 117, \dodoi{10.3847/0004-637X/824/2/117}

\bibitem[{{Ram{\'\i}rez} {et~al.}(2012){Ram{\'\i}rez}, {Fish}, {Lambert}, \&
  {Allende Prieto}}]{Ramirez2012}
{Ram{\'\i}rez}, I., {Fish}, J.~R., {Lambert}, D.~L., \& {Allende Prieto}, C.
  2012, \apj, 756, 46, \dodoi{10.1088/0004-637X/756/1/46}

\bibitem[{{Scholz}(2016)}]{Scholz2016}
{Scholz}, R.~D. 2016, \aap, 587, A51, \dodoi{10.1051/0004-6361/201527965}

\bibitem[{{Spina} {et~al.}(2018){Spina}, {Mel{\'e}ndez}, {Karakas}, {dos
  Santos}, {Bedell}, {Asplund}, {Ram{\'\i}rez}, {Yong}, {Alves-Brito}, {Bean},
  \& {Dreizler}}]{Spina2018}
{Spina}, L., {Mel{\'e}ndez}, J., {Karakas}, A.~I., {et~al.} 2018, \mnras, 474,
  2580, \dodoi{10.1093/mnras/stx2938}

\bibitem[{{Stephens} {et~al.}(2009){Stephens}, {Leggett}, {Cushing}, {Marley},
  {Saumon}, {Geballe}, {Golimowski}, {Fan}, \& {Noll}}]{Stephens2009}
{Stephens}, D.~C., {Leggett}, S.~K., {Cushing}, M.~C., {et~al.} 2009, \apj,
  702, 154, \dodoi{10.1088/0004-637X/702/1/154}

\bibitem[{{Takeda} {et~al.}(2007){Takeda}, {Ford}, {Sills}, {Rasio}, {Fischer},
  \& {Valenti}}]{Takeda2007}
{Takeda}, G., {Ford}, E.~B., {Sills}, A., {et~al.} 2007, \apjs, 168, 297,
  \dodoi{10.1086/509763}

\bibitem[{{Thalmann} {et~al.}(2009){Thalmann}, {Carson}, {Janson}, {Goto},
  {McElwain}, {Egner}, {Feldt}, {Hashimoto}, {Hayano}, {Henning}, {Hodapp},
  {Kandori}, {Klahr}, {Kudo}, {Kusakabe}, {Mordasini}, {Morino}, {Suto},
  {Suzuki}, \& {Tamura}}]{Thalmann2009}
{Thalmann}, C., {Carson}, J., {Janson}, M., {et~al.} 2009, \apjl, 707, L123,
  \dodoi{10.1088/0004-637X/707/2/L123}

\bibitem[{{Zurlo} {et~al.}(2016){Zurlo}, {Vigan}, {Galicher}, {Maire}, {Mesa},
  {Gratton}, {Chauvin}, {Kasper}, {Moutou}, {Bonnefoy}, {Desidera}, {Abe},
  {Apai}, {Baruffolo}, {Baudoz}, {Baudrand}, {Beuzit}, {Blancard},
  {Boccaletti}, {Cantalloube}, {Carle}, {Cascone}, {Charton}, {Claudi},
  {Costille}, {de Caprio}, {Dohlen}, {Dominik}, {Fantinel}, {Feautrier},
  {Feldt}, {Fusco}, {Gigan}, {Girard}, {Gisler}, {Gluck}, {Gry}, {Henning},
  {Hugot}, {Janson}, {Jaquet}, {Lagrange}, {Langlois}, {Llored}, {Madec},
  {Magnard}, {Martinez}, {Maurel}, {Mawet}, {Meyer}, {Milli},
  {Moeller-Nilsson}, {Mouillet}, {Orign{\'e}}, {Pavlov}, {Petit}, {Puget},
  {Quanz}, {Rabou}, {Ramos}, {Rousset}, {Roux}, {Salasnich}, {Salter},
  {Sauvage}, {Schmid}, {Soenke}, {Stadler}, {Suarez}, {Turatto}, {Udry},
  {Vakili}, {Wahhaj}, {Wildi}, \& {Antichi}}]{Zurlo2016}
{Zurlo}, A., {Vigan}, A., {Galicher}, R., {et~al.} 2016, \aap, 587, A57,
  \dodoi{10.1051/0004-6361/201526835}

\end{thebibliography}


\end{document}